\documentclass[preprint,review,12pt]{elsarticle}

\usepackage{amssymb}
\usepackage{amsmath}

\usepackage{graphicx}
\usepackage{subcaption}
\usepackage{multirow}
\usepackage{booktabs}
\usepackage{mathtools}
\usepackage{url}

\journal{Energy Storage}

\begin{document}

\begin{frontmatter}



\title{Novel Low-Complexity Model Development for Li-ion Cells Using Online Impedance Measurement}


\author{Abhijit Kulkarni, Ahsan Nadeem, Roberta Di Fonso, Yusheng Zheng, and Remus Teodorescu} 

\affiliation{organization={Aalborg University},
            addressline={Department of Energy}, 
            city={Aalborg},
            postcode={9220}, 
            country={Denmark}}


\begin{abstract}
Modeling of Li-ion cells is used in battery management systems (BMS) to determine  key states such as state-of-charge (SoC), state-of-health (SoH), etc. Accurate models are also useful in developing a cell-level digital-twin that can be used for protection and diagnostics in the BMS. 
In this paper, a low-complexity model development is proposed based on the equivalent circuit model (ECM) of the Li-ion cells. The proposed approach uses online impedance measurement at discrete frequencies to derive the ECM that matches closely with the results from the electro-impedance spectroscopy (EIS). The proposed method is suitable to be implemented in a microcontroller with low-computational power, typically used in BMS. Practical design guidelines are proposed to ensure fast and accurate model development. 
Using the proposed method to enhance the functions of a typical automotive BMS is described.  Experimental validation is performed using large prismatic cells and small-capacity cylindrical cells. Root-mean-square error (RMSE) of less than 3\%  is observed for a wide variation of operating conditions. 
\end{abstract}



\end{frontmatter}



\section{Introduction}
\label{sec1}
The demand for Li-ion cells is growing rapidly due to the increased penetration of electric vehicles (EVs) in the transportation sector \cite{lib_demand_elsv}. There is also an active development of electric vertical takeoff and landing aircraft (eVTOL) for the applications such as air-taxis and air-ambulances. The general passenger aerospace industry is also expected to incorporate more electric aircraft (MEA) in increasing numbers \cite{lib_aero_demand_elsv}. Thus, the demand of Li-ion cells is also leading to the demand of next generation battery management systems (BMSs) that enable very high levels of safety and increase the operating life of the batteries used within the e-transportation application \cite{bms_high_level}.

Typical BMS used in EVs perform the functions of cell voltage measurement, temperature monitoring, state estimation (SoC, SOH), cell balancing, isolation sensing and protection \cite{bms_high_level,plett_vol1}. The state estimation is done using the sensed signals and model of the Li ion cells and packs. Typically equivalent circuit based models are used since they are computationally less intensive compared to the physics based models such as P2D model \cite{model_complexity_good}. Recently, there is a lot of research focusing on using artificial intelligence (AI) based methods for improving the accuracy of the state estimations, especially the SoC and SoH \cite{soc_ai_tpel,soc_ai_elsv1,soc_ai_tii1,soc_soh_ai_access,aau_perspective, soh_review_elsv1,ai_soh_elsv1}. 

\begin{figure}[h]
  \centering
  \begin{subfigure}[b]{0.49\textwidth}
    \includegraphics[width=\textwidth]{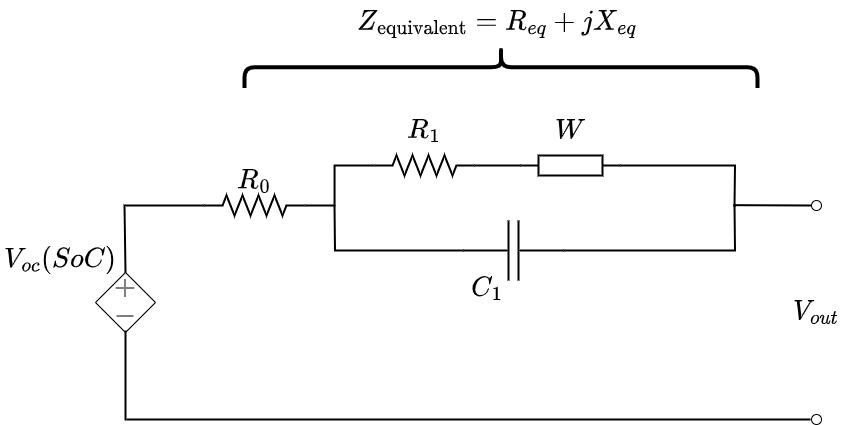}
    \caption{Randles circuit based ECM.}
  \end{subfigure}
  \hfill
  \begin{subfigure}[b]{0.49\textwidth}
    \includegraphics[width=\textwidth]{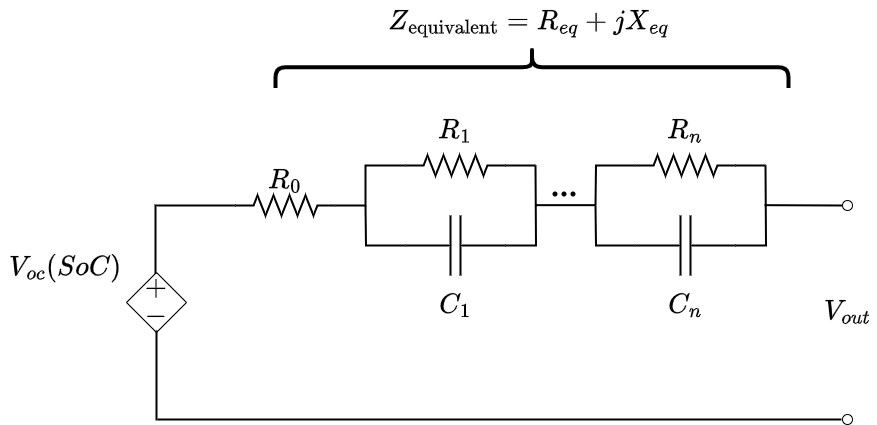}
    \caption{Multiple RC branch-based ECM.}
  \end{subfigure}
  \caption{Popular equivalent circuit models used in Li-ion cells.}
  \label{fig:ecm_two_types}
\end{figure}

BMS can be added with additional functions to enable enhanced protection, monitoring and diagnostics. One approach to achieve this is to implement a cell-level digital-twin \cite{cell_level_twin,sb_cell_level_twin,quasi_cell_level_twin,roberta_twin,cell_level_twin_mdpi_good}. 
The digital-twin can be tailored to perform a specific function within the BMS. One function of interest is estimating the SoH. 
Considering that the controller used in BMS may not be with high computational power due to cost-limitations, it is preferable to have a digital-twin that can be implemented using a simple model. Thus, ECM are attractive since they essentially involve a state-space model whose parameters change depending on the operating conditions of the cell \cite{cell_level_twin_mdpi_good}. Fig. 1 shows two types of ECM that are commonly used \cite{plett_vol1}. Fig. \ref{fig:ecm_two_types} (a) shows Randles circuit based ECM and Fig. \ref{fig:ecm_two_types}(b) shows the ECM with n-number of RC branches that replace the Warburg element used in Randles circuit.
The accuracy of the model with multiple RC branches increases when more branches are used. However, it is reported that 3RC branches may be sufficient to capture the desired characteristics from the model \cite{3rc_greg}.
Note that some researchers use ECMs with constant phase elements (CPE) in place of the capacitances \cite{cpe_elsv}. 

The ECM of Li-ion cells represents the variation in the output voltage for given load and operating conditions. Offline techniques such as EIS or hybrid pulse power characteristic (HPPC) are normally used to derive the parameters of the ECM such as the values of resistances, capacitances, Warburg element, etc for varying operating conditions such as temperature, SoC and SoH. This is usually not feasible to implement online since the procedure takes a long time that can impact the operation of the battery pack. Also, EIS methods typically use small-signal current disturbances in terms of varying sinuoids. Thus, in-field EIS with higher current is not same as the classical offline EIS. This is termed as operando EIS \cite{operando_eis}. The trends of classical and operando EIS are similar and hence it can be used for understanding the behavior of a cell during operation.
Even if the operando impedance is measured online, it needs to be fit into the ECM parameters to make useful predictions. 
Conventional methods use optimization techniques to fit the model parameters, which is a computationally intensive effort. It is difficult to implement in embedded microcontrollers. On the other hand, time-domain based methods such as HPPC need accurate timing information and will require a long time to provide the settling time to compute the capacitances of the ECM. 

In this paper, a new low-complexity method is proposed to identify the ECM parameters online. This method is based on the use of three discrete impedance values at three different frequencies. Analytical equations are derived that provide closed-form expressions to the ECM parameters such as the resistances, capacitance and Warburg element gain. This paper also describes how to choose the three frequencies and shows the impact of the selection on the model accuracy. Online evaluation of these parameters in a  battery management system is discussed. The proposed approach does not involve any online optimization or the use of complex curve-fitting algorithms. Hence, it is computationally very simple and can be implemented in low-cost microcontrollers that are used in the BMS. All the analytical equations and the proposed methodology has been validated experimentally using two types of cells, namely, 50Ah Li-ion cells of prismatic geometry and 4.85Ah cylindrical cells. The model thus developed can be used as a digital-twin as an extension and it is useful for accurate state estimations such as state-of-temperature, SoH, etc.

This paper is organized as follows. Section 2 describes the ECM considered in this paper and the variation of the parameters of the ECM with respect to the operating condition of the cell. In Section 3, analytical expressions are derived that are used to develop the ECM using only 3 discrete impedance values. Practical implementation of the proposed method in  is discussed in Section 4. Experimental setup and validation are detailed in Section 5 followed by conclusions in Section 6.

\section{Overview of the Equivalent Circuit Model of Li-ion Cells}
Li-ion cell is a complex source of electrical energy whose terminal voltage is a result of the chemical reactions within the cell. The terminal voltage varies mainly depending on the SoC, temperature, and SoH. Most of the models developed have different advantages/ limitations and may apply well in certain scenarios than others. For example, ECMs are very good in quickly predicting the terminal voltage at a given instant of time and their parameters can be related to the internal electrochemical reactions inside the cell. 
The physics based models are well-suited for capturing the internal chemical reactions, electrode voltages and degradation. However, they suffer from the disadvantage of having large number of parameters and high computational burden due to the use of partial differential equations.

The parameters of the Randles circuit shown in Fig.\ref{fig:ecm_two_types}(a) can be connected to the internal behaviors of a cell \cite{plett_vol1}. For example, the resistance $R_0$ models the resistance of the electrolyte. $R_1$ represents the voltage drop across the electrode and electrolyte interface and it is also termed as charge-transfer resistance. $C_1$ is called the double layer capacitance and it models the build up of charges on the electrode surface. The last parameter $W$ is the Warburg element and it represents the Li ion diffusion in the electrodes. Note that the ECM is composed of linear circuit elements except the Warburg element. It is represented as an impedance that is inversely proportional to the square-root of the frequency and provides always a phase lag of $45^\circ$. Eqn. (\ref{eq:def_warburg}) shows the expression for Warburg element.
\begin{equation}
    W = \frac{A_w}{\sqrt{j\omega}} \label{eq:def_warburg}
\end{equation}

\begin{figure}
    \centering
    \includegraphics[width=0.75\linewidth]{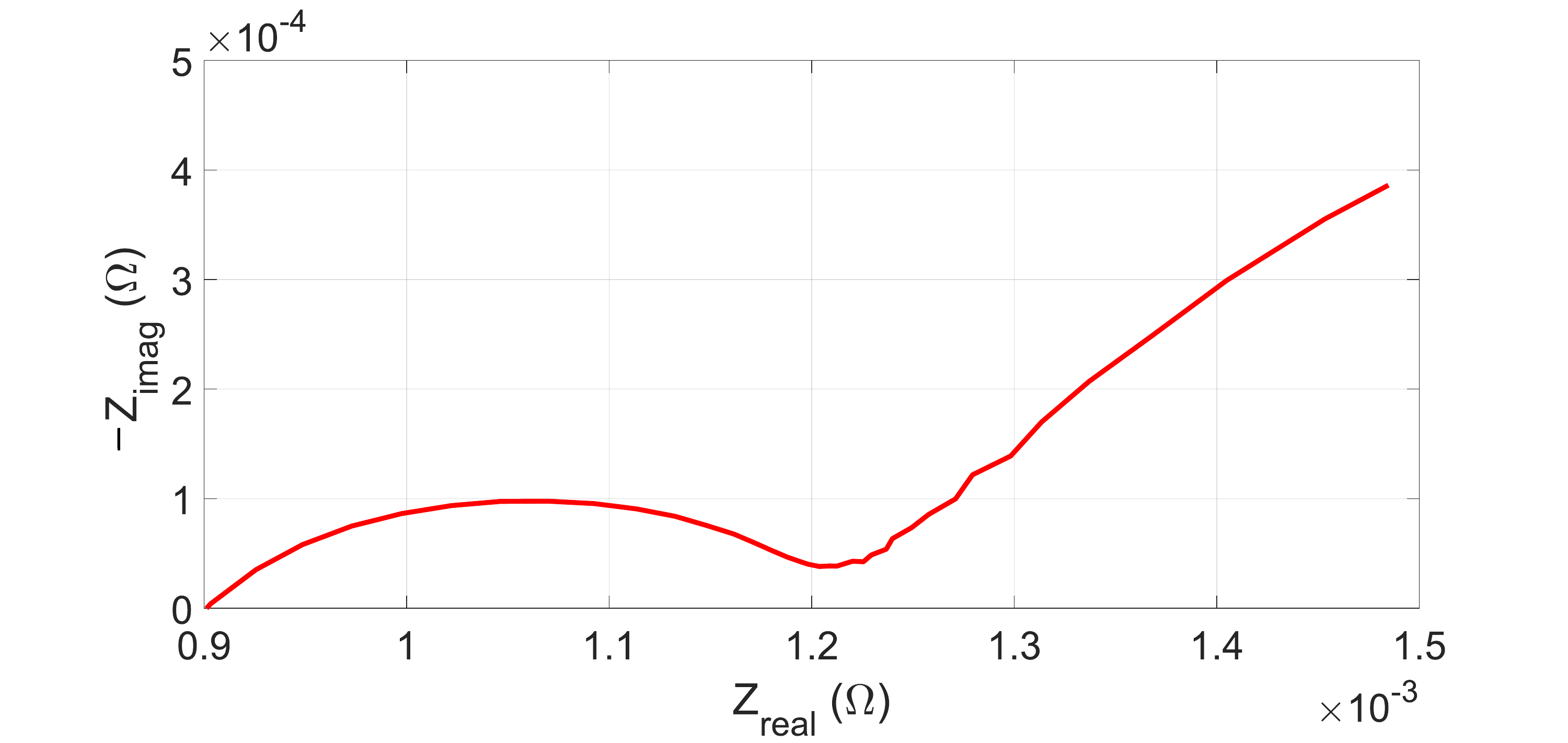}
    \caption{Nyquist plot of a 50Ah Li ion cell at 50\% SoC and at a temperature of $25^oC$.}
    \label{fig:nyquist_ex}
\end{figure}

The ECM with Warburg element represents the frequency response of Li-ion cells with high accuracy. Typical Nyquist plot is shown in Fig. \ref{fig:nyquist_ex}. This is for a 50Ah prismatic cell with nickel manganese cobalt oxide (NMC) chemistry at 50\% SoC and at a temperature of $25^\circ C$. The lowest frequency point is located at the top right of the curve and frequency increases as the Nyquist plot is traced from top right to bottom left direction. Warburg element is responsible for the linear rise of the impedance for lower frequencies. The semicircular part is due to the RC branch when the impact of the Warburg element is negligible. At higher frequencies, the impedance essentially reduces to electrolytic resistance $R_0$. At very high frequencies, parasitic inductances play a role (not shown in Fig. \ref{fig:ecm_two_types}) and appear to increase the impedance magnitude with frequency.

\subsection{Variation of ECM parameters with battery states}
It is well known that the impedance of the cell is impacted by the temperature, SoC, and SoH. Typical variations observed for these parameters are summarized in Table \ref{tab:ecm_var_summary}.

\begin{table}[ht]
\small{
\centering
\caption{Typical variation of Randles circuit parameters.}
\begin{small}
\begin{tabular}{|c|c|c|c|}
\hline
\multirow{2}{*}{Parameter} & \multirow{2}{*}{Variation with} &\multirow{2}{*}{Variation with} & \multirow{2}{*}{Variation with} \\  
& increasing temperature & increasing SoC & decreasing SoH \\
\hline
$R_0$ &  $\downarrow$ &  $\downarrow$ &  $\uparrow$ \\
\hline
$R_1$ &  $\downarrow$ & Non-monotonic &  $\uparrow$ \\
\hline
$C_1$ &  $\uparrow$ & Non-monotonic &  $\uparrow$ \\
\hline
$A_w$ &  $\downarrow$ & Non-monotonic & $\uparrow$ \\
\hline
\end{tabular}
\end{small}
\label{tab:ecm_var_summary}}
\end{table}

\begin{figure}
  \centering
  \begin{subfigure}[b]{0.495\textwidth}
    \includegraphics[width=\textwidth]{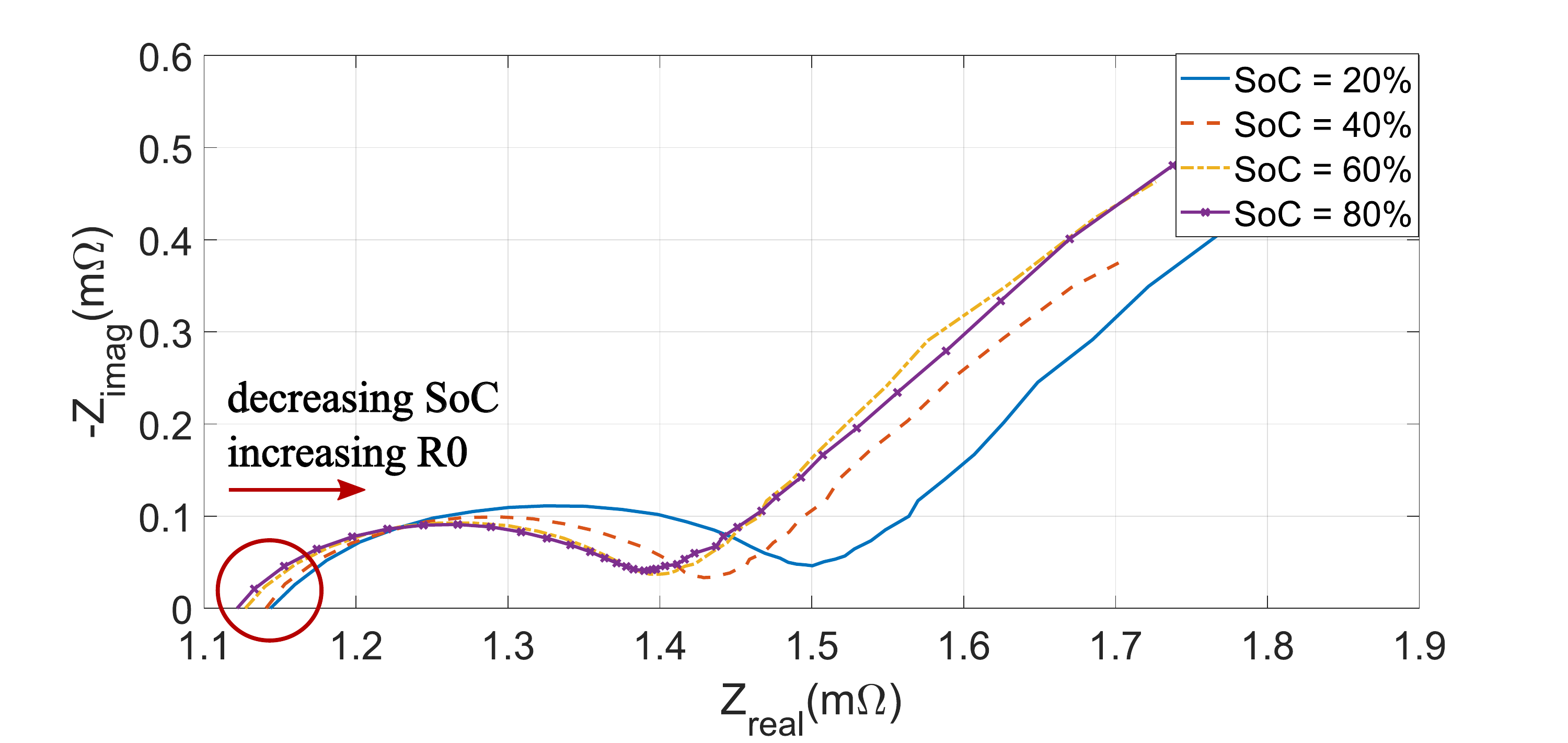}
    \caption{Nyquist plots at different SoC but at a temperature of $25^oC$.}
  \end{subfigure}
  \hfill
  \begin{subfigure}[b]{0.495\textwidth}
    \includegraphics[width=\textwidth]{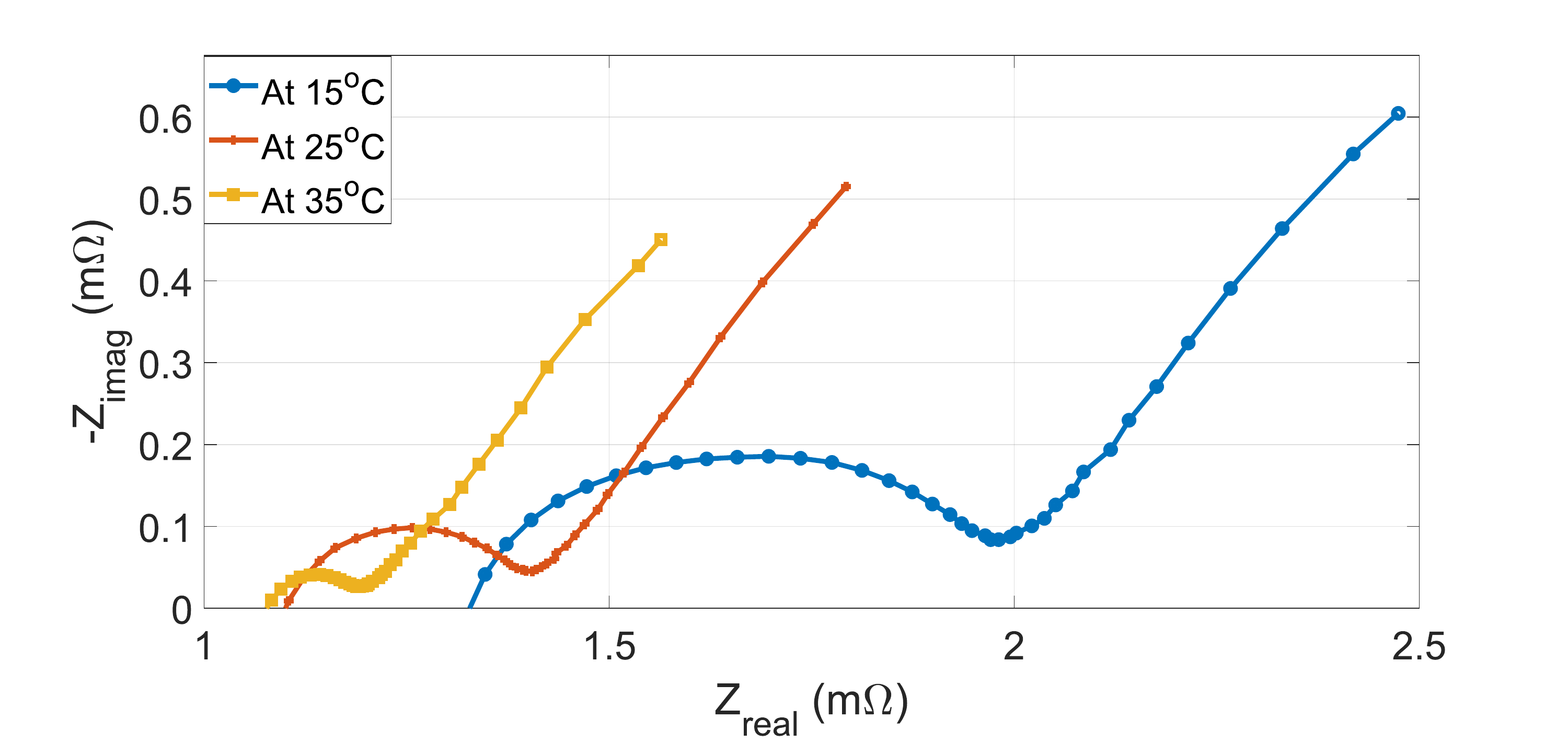}
    \caption{Nyquist plots at different temperatures but at a constant $SoC = 50\%$.}
  \end{subfigure}
  \caption{Illustrating the impact of SoC and Temperature on the battery impedance using Nyquist plots.}
  \label{fig:nyquist_variation_lab}
\end{figure}

Fig. \ref{fig:nyquist_variation_lab} shows the parameter variations with respect to SoC and temperature. This is based on EIS tests.
It can be seen that the variation summarized in Table \ref{tab:ecm_var_summary} can be visualized using Fig. \ref{fig:nyquist_variation_lab} for both SoC and temperature variation. The variation is clearly observable for the electrolytic resistance $R_0$, which is the x-intercept, it increases with lower SoC and lower temperature. However, the variation of the other parameters is known and can be further evaluated based on the proposed analytical solution discussed in Section 3.
Similar results for the impedance variation are summarized well in \cite{impedance_impact_curves,eis_temp_loughborough}.

\subsection{Review of the existing methods to obtain ECM parameters}
Typically, EIS is used to obtain the cell impedances for a wide range of frequencies and an optimization method is used to fit the ECM parameters to the results from EIS. This is an accurate approach for offline analysis. It can be used to study the impact of operating conditions on the ECM parameters. However, it is not suitable for online method since EIS cannot be performed online. There are other approaches that use pseudo-random binary sequences (PRBS) \cite{prbs_elsv1,binary_seq_ies} instead of multi-sine EIS. Using binary sequences helps in reducing the time for the test compared to classical EIS. However, these methods use Fourier analysis to compute the impedance at different frequency components making them computationally intensive. Additionally, it is impacted by poor signal to noise ratio \cite{prbs_limitation_ecce} and may require additional hardware and control platform to produce these signals.


This paper tries to address these deficiencies by enabling online measurement of impedance using a smart BMS and a low-complexity analytical method to obtain the ECM parameters using just three discrete impedance values. This is described in detail in Section 3.

\section{Proposed Model Development Using Discrete Impedance Values}
\begin{figure}
    \centering
    \includegraphics[width=0.6\linewidth]{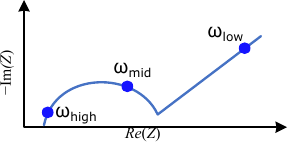}
    \caption{Discrete frequencies marked on a generic Nyquist plot to obtain the ECM parameters analytically.}
    \label{fig:discrete_freq_marked}
\end{figure}

In Fig. \ref{fig:discrete_freq_marked}, the basic philosophy of the proposed approach is shown. There are three discrete frequencies, $\omega _{low}$, $\omega _{mid}$ and $\omega _{high}$, that are chosen. Then analytical expressions for the impedance are derived based on the Randles circuit. These expressions are manipulated algebraically to obtain the values of the resistors, capacitors and Warburg element. Note that the selection of frequencies is critical and it impacts the accuracy of the results. 
Consider a high frequency impedance, where the frequency considered is termed as $\omega _{high}$. The value of the high frequency depends on the type of the cell. It should also be low enough to avoid the impacts of stray inductances. In Section 3.1, an example calculation is demonstrated showing the frequency selection.
For the frequency $\omega _{high}$, the capacitive impedance and the Warburg impedance will be negligibly small, hence, the equivalent impedance is given by,
\begin{equation}
    Z_{eq}(j\omega _{high}) \approx R_0 \label{eq:R0_expr}
\end{equation}
Now, consider the case of low frequency $\omega _{low}$ marked on Fig. \ref{fig:discrete_freq_marked}. This falls under the influence of Warburg element. This is also evident from the ECM, where the capacitive impedance can be considered high enough, so the parallel branch impedance will be dominated by the Warburg impedance. The equivalent impedance for this frequency from the ECM is given by,
\begin{equation}
     Z_{eq}(j\omega _{low}) \triangleq Z_{low} \approx R_0 + R_1 + \frac{A_w}{\sqrt{j\omega_{low}}} \label{eq:zlow_main_eq}
\end{equation}
The real and imaginary parts of Eq. \ref{eq:zlow_main_eq} can be separted algebraically as,
\begin{align}
    Z_{low,real} &= R_0 + R_1 + \frac{A_w}{\sqrt{2\omega_{low}}} \label{eq:zlowreal} \\
    Z_{low,imag} &= - \frac{A_w}{\sqrt{2\omega_{low}}} \label{eq:zlowimag} 
\end{align}
The real and imaginary parts of the impedance are assumed to be known. For offline processes, EIS can provide the values. The online process described in Section 4 also provides these values. Based on this information, Eq. \ref{eq:zlowimag} can be rearranged to obtain,
\begin{equation}
    A_w = |Z_{low,imag}|\sqrt{2\omega_{low}} \label{eq:aw_expr}
\end{equation}
Using Eq. \ref{eq:aw_expr} and Eq. \ref{eq:R0_expr} in Eq. \ref{eq:zlowreal}, the resistance $R_1$ can be determined as,
\begin{equation}
    R_1 = Z_{low,real} - R_0 - \frac{A_w}{\sqrt{2\omega_{low}}} \label{eq:R1expr}
\end{equation}
Now, out of the 4 parameters of the ECM, three have been determined. The last parameter is the capacitance $C_1$. This is computed using the impedance at a mid-frequency range, where the capacitance will have an appreciable impact. Here, the impedance due to Warburg element is negligible, and can be ignored. The Randles circuit will reduce to just the parallel combination between $R_1$ and $C_1$ in this frequency range. With this assumption, the impedance at the mid-frequency range can be expressed as,
\begin{equation}
     Z_{eq}(j\omega _{mid}) \triangleq Z_{mid} \approx R_0 + \frac{R_1}{1+j\omega_{mid} R_1C_1} \label{eq:zmid_main_eq}
\end{equation}
This can be separated into real and imaginary parts as,
\begin{align}
    Z_{mid,real} &= R_0 + \alpha \label{eq: zmid_real} \\
    Z_{mid,imag} &= \alpha \omega_{mid} R_1C_1 \label{eq: zmid_imag}
\end{align}
where,
\begin{equation}
    \alpha = \frac{R_1}{1+(\omega_{mid}R_1C_1)^2} \label{eq:alpha_wmid}
\end{equation}
Using the real component of the impedance, the term $\alpha$ can be computed using Eq. \ref{eq: zmid_real} since the real part of the impedance at this frequency is already known either from offline or online measurement and $R_0$ is determined using Eq. \ref{eq:R0_expr}. 

The capacitance value is determined simply by rearranging Eq. \ref{eq: zmid_imag},
\begin{equation}
    C_1 = \frac{Z_{mid,imag}}{\alpha \omega_{mid} R_1} \label{eq: C1_expr}
\end{equation}

Thus the set of equations (\ref{eq:R0_expr}), (\ref{eq:aw_expr}), (\ref{eq:R1expr}) and (\ref{eq: C1_expr}) are used to determine all the parameters of the ECM in Fig. \ref{fig:ecm_two_types}(a) based on only three discrete impedance values. It is reiterated again that the impedance values can be obtained offline using a traditional method or online as discussed in Section 4.  

\subsection{Frequency selection approach}
Commercial EIS equipment use a frequency range of few millihertz to tens of kilohertz. Impedance is evaluated at a large number of points within this range to result in the typical Nyquist or Bode plots. Cell capacity and geometry determine the maximum relevant frequency. For example, 50 Ah prismatic cell used in this work needs a maximum frequency of about 500 Hz. Beyond this, the impedance is mainly the electrolytic resistance $R_0$ and  stray inductance. The dominance of the inductance can be observed when the impedance starts increasing with frequency in the high frequency domain.

The selection of minimum frequency ($\omega_{low}$) is impacted by two factors: first is the accuracy in estimating the ECM parameters and the second one is the time taken to develop the model. For example, $\omega_{low}$ corresponding to 0.01 Hz, needs at least 3 cycles to obtain the impedance. Thus, it will take $\geq 300~s$ or 5 minutes to get the impedance value. On the otherhand, a higher value of $\omega_{low}$, say 10 Hz, will give a fast result but will significantly impact the accuracy since at that frequency the effect of double-layer capacitance $C_1$ cannot be ignored. The selection of frequencies for  experimental data is illustrated with an example in Section 5.1.

\begin{figure}
    \centering
    \includegraphics[width=0.85\linewidth]{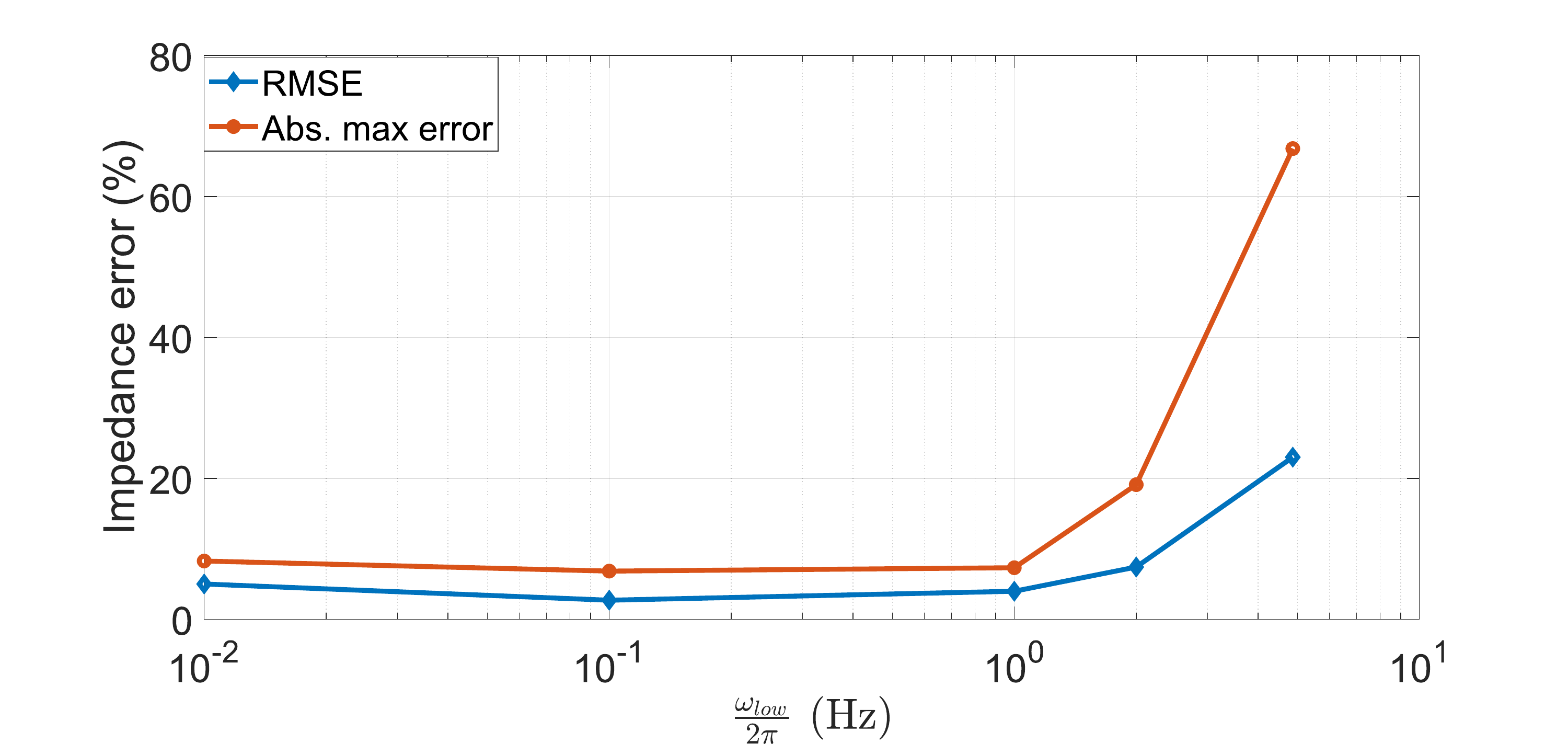}
    \caption{Impact of frequency selection on the model accuracy.}
    \label{fig:wlow_selection}
\end{figure}

Fig. \ref{fig:wlow_selection} shows the variation of the root-mean-square error (RMSE) and absolute maximum error (AME) between the impedances measured from the model and from EIS for 5 different selections of $\omega_{low}$. As it can be observed, up to 1Hz, the RMSE and AME are practically constant. 
Hence, considering the fast response time for the impedance computation, a selection of 1 Hz seems appropriate. However, around the frequency of 0.1Hz, RMSE is minimum. The time required for impedance estimation at 0.1 Hz will be about 30 s, which is considerably shorter than offline EIS based method. The mid-frequency $\omega_{mid}$ is to be chosen in the earlier parts of the semicircular area on the Nyquist plot as indicated in Fig. \ref{fig:discrete_freq_marked}. Note that this frequency does not significantly impact the computation time since it is in the order of few tens of hertz.

\section{Using Proposed Method with Online Impedance Measurement}

\begin{figure}[htbp]
    \centering
    \includegraphics[width=0.75\linewidth]{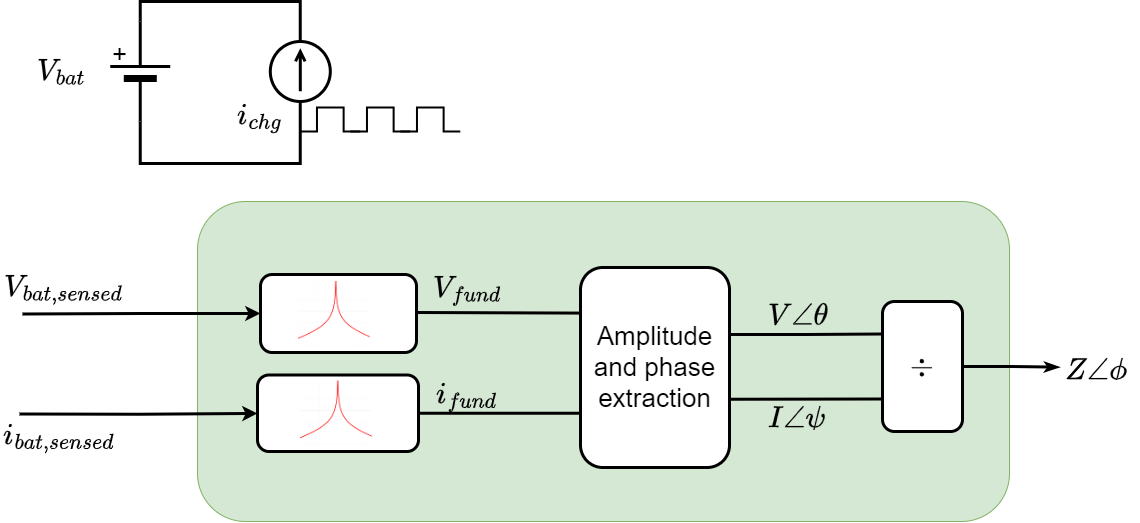}
    \caption{Proposed online impedance extraction method using pulsed current.}
    \label{fig:online_imp_calc}
\end{figure}

The proposed method provides a simplified solution for obtaining the ECM parameters online. Instead of applying wideband signals, pulsed current for a certain duration is applied. Three different pulsed currents are necessary as per the method described in Section 3. Since single-frequency current is applied, Fourier analysis is not necessary for post processing. Simple filters can be used to extract the fundamental components of the cell voltage and current. Amplitude and phase can be extracted from the fundamental components to estimate the impedance at the given frequency. This process is repeated for the three-frequencies and the algebraic equations derived in Section 3 are used to obtain the ECM parameters. Thus, the proposed approach uses only simple filters and algebraic equations to get the ECM, which results in fast development of the model parameters even in microcontrollers with low computation power. Fig. \ref{fig:online_imp_calc} shows the proposed approach.

Note that if conventional battery packs and BMS are used, a charger capable of producing pulsed current will be needed. In such systems, the model developed can be used to understand the aging of cells. However, there is a lot of research on improvising the conventional BMS. In \cite{aau_perspective}, a smart BMS is proposed where each cell has an integrated half-bridge. This architecture can be used to produce the pulsed current at cell level during any operational mode - charging or discharging. In such a system, it is possible to estimate the model parameters whenever needed and can be used for various diagnostic functions such as sensorless temperature estimation \cite{yusheng}.

\subsection{Filter design}
The impedance estimation method shown in Fig. \ref{fig:online_imp_calc} uses two filters to extract the fundamental component in the cell voltage and current. In the simplest implementation, the filter is a bandpass filter with the following transfer function.

\begin{equation}
    G(s) = \frac{ks\omega_0}{s^2+ks\omega_0+\omega^2_0} \label{eq:filt_tf}
\end{equation}

\begin{figure}
  \centering
  \begin{subfigure}[b]{0.495\textwidth}
    \includegraphics[width=\textwidth]{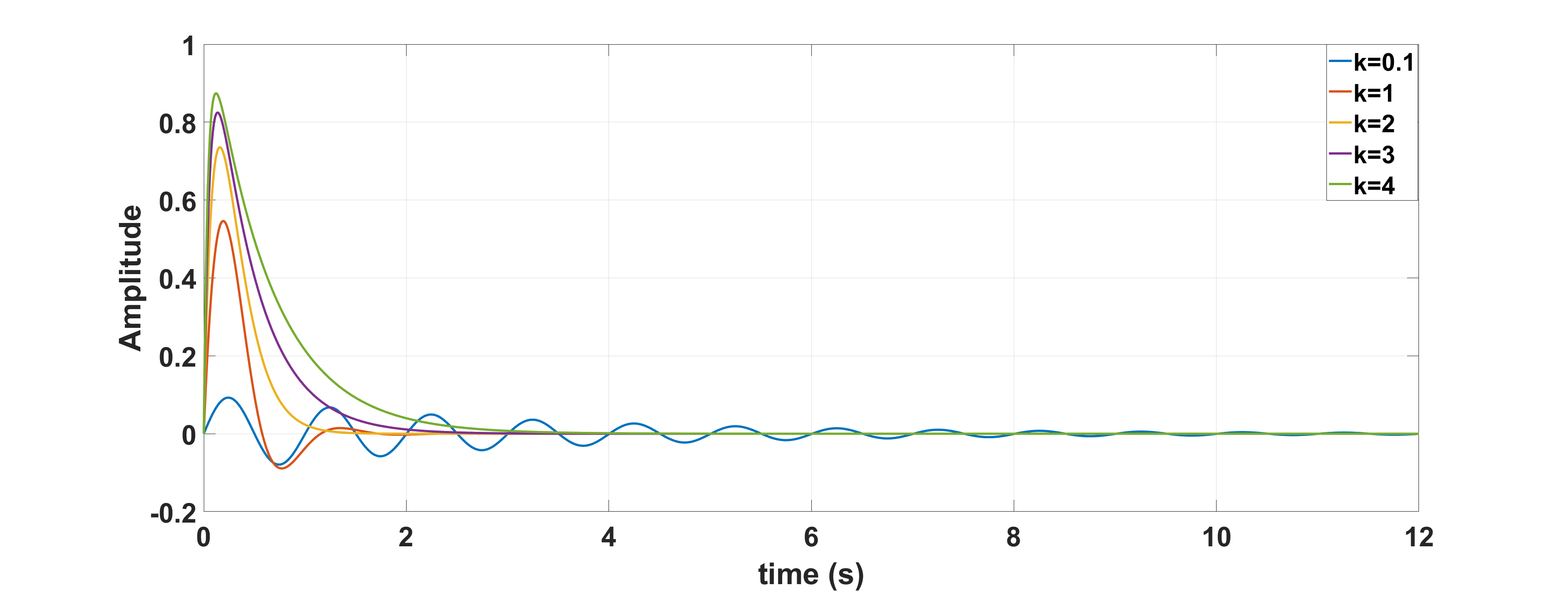}
    \caption{Step response.}
  \end{subfigure}
  \hfill
  \begin{subfigure}[b]{0.495\textwidth}
    \includegraphics[width=\textwidth]{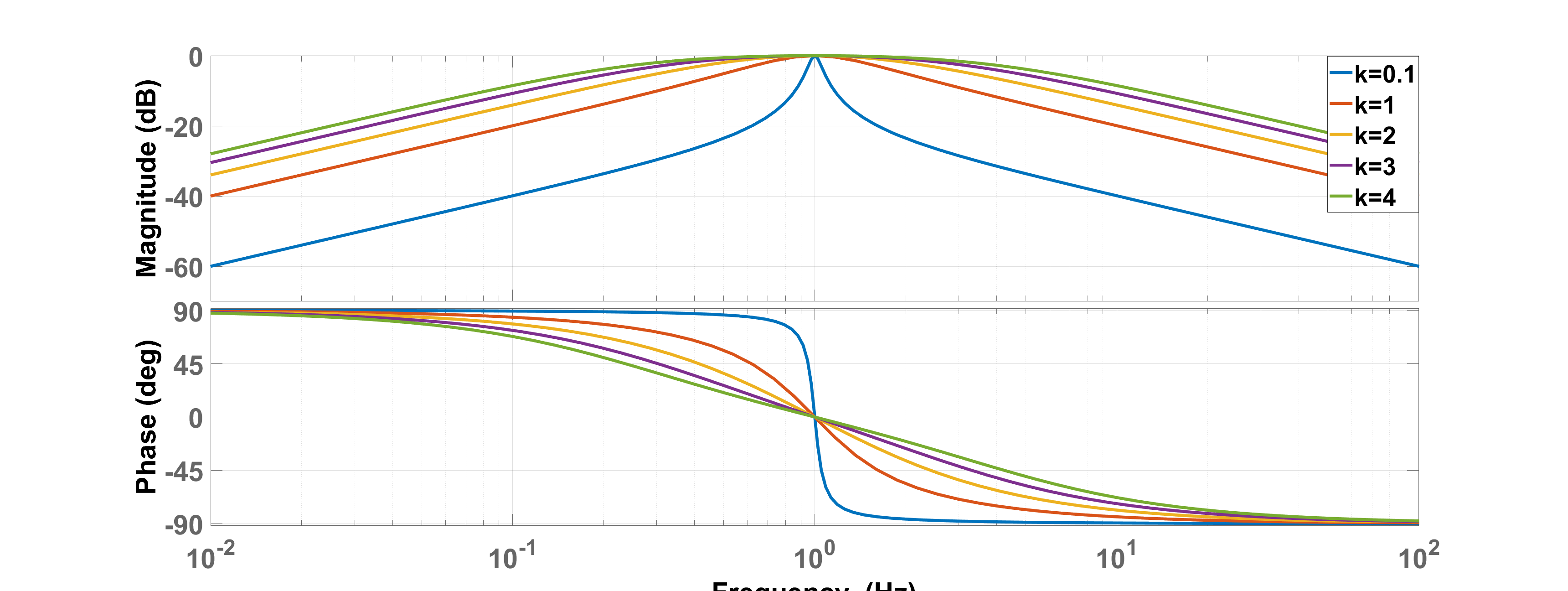}
    \caption{Bode plot.}
  \end{subfigure}
  \caption{Response of the band-pass filter for varying $k$ and $f_0= 1Hz$.}
  \label{fig:filt_resp}
\end{figure}

The filter transfer function in Eq. \ref{eq:filt_tf} represents a simple second-order band-pass filter (BPF). The term $\omega_0 = 2\pi f_0$ is chosen based on the pulse frequency. As per the analytical equations in Section 3, this can be either $\omega_{low}$ or $\omega_{mid}$ or $\omega_{high}$. The other parameter in the transfer function is $k$. This affects the attenuation of the filter around the center point $\omega_{0}$ and the settling time as shown in Fig. \ref{fig:filt_resp}. A value of $k=1$ gives a good compromise between the attenuation and settling time and is used in this work. This filter can be cascaded twice to result in a fourth-order filter if higher harmonic attenuation is needed from sensed signals.

\subsection{Applying the proposed method in BMS}
\begin{figure}[h]
    \centering
    \includegraphics[width=0.75\linewidth]{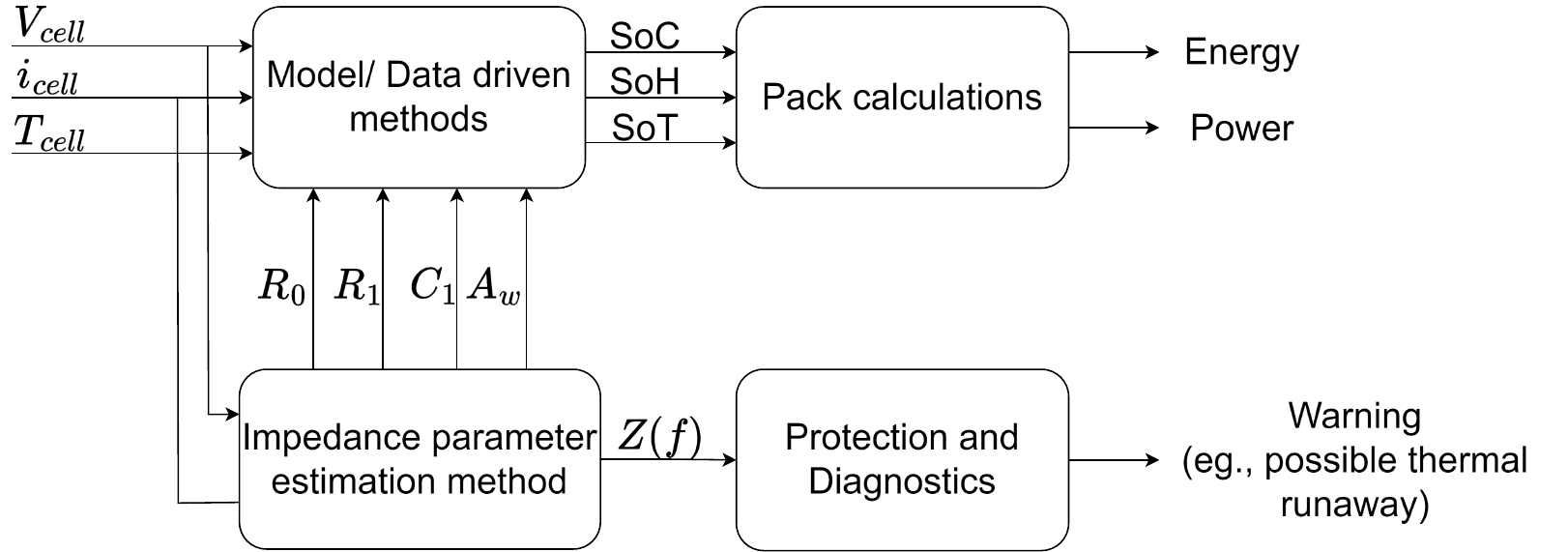}
    \caption{Enhancing the BMS for e-transport applications using the proposed method.}
    \label{fig:new_bms}
\end{figure}

The proposed ECM parameter extraction method using pulsed signals can be applied in a modern BMS system for automotive applications. Conventional BMS use the cell voltage, current and temperature inputs to determine SoC, SOH and SoT at cell-level along with pack-level power and energy \cite{plett_vol1}. The proposed method determines the ECM parameters that can be input to the cell-level state calculation methods, which can be model-based or data-driven. The ECM parameters can act as additional features for data-driven methods for enhancing the estimation accuracy. The complete system is shown in Fig. \ref{fig:new_bms}. The ECM parameters or impedance value can be used for protection and diagnostic part of the BMS. For example, impedance at a certain frequency ($Z(f)$) can be used as an indicator of cell temperature \cite{yusheng}. Thus, it can be used as a measure to provide warning for fast temperature rise/ possible thermal runaway. This will be an additional useful information along with the sensors and can enhance the system safety and reliability with redundancy. 

\section{Experimental Validation}
There are two concepts proposed in this paper that are experimentally validated. These are described in the following subsections.

\subsection{Validating the proposed ECM parameter identification approach} An NMC cell (50Ah) \cite{calb} is tested in an EIS instrument from Digatron at various SoC and temperature conditions. The output of the test is a list of impedance magnitude and phase for a wide frequency range as provided by the EIS instrument. One sample Nyquist plot from this test is shown in Fig.\ref{fig:nyquist_ex}.

\begin{figure}[h]
    \centering
    \includegraphics[width=0.85\linewidth]{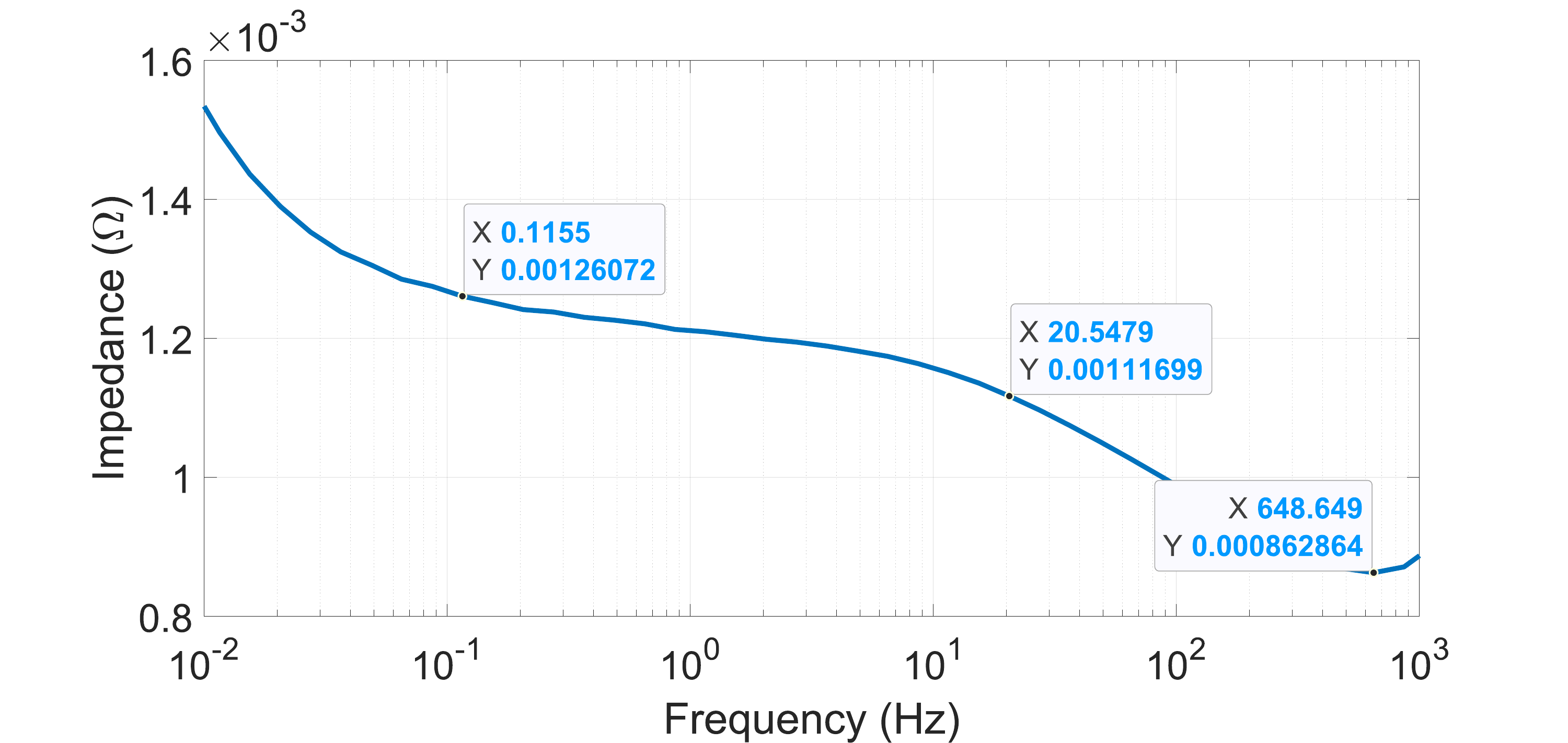}
    \caption{EIS impedance curve of a 50Ah cell with the 3 impedance points marked.}
    \label{fig:mag_plot_freq_marked}
\end{figure}

\begin{figure}
  \centering
  \begin{subfigure}[b]{0.495\textwidth}
    \includegraphics[width=\textwidth]{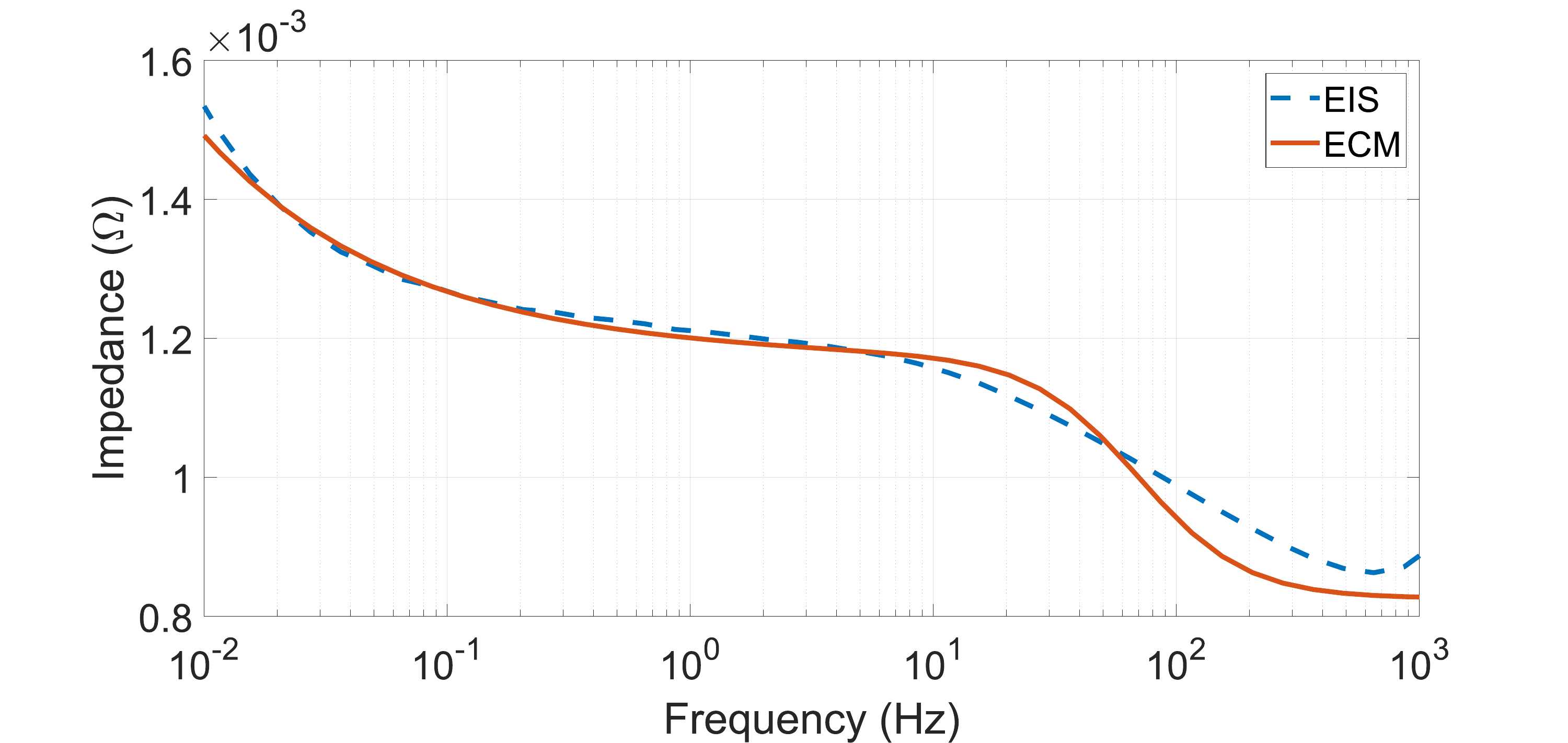}
    \caption{Impedance magnitude plot.}
  \end{subfigure}
  \hfill
  \begin{subfigure}[b]{0.495\textwidth}
    \includegraphics[width=\textwidth]{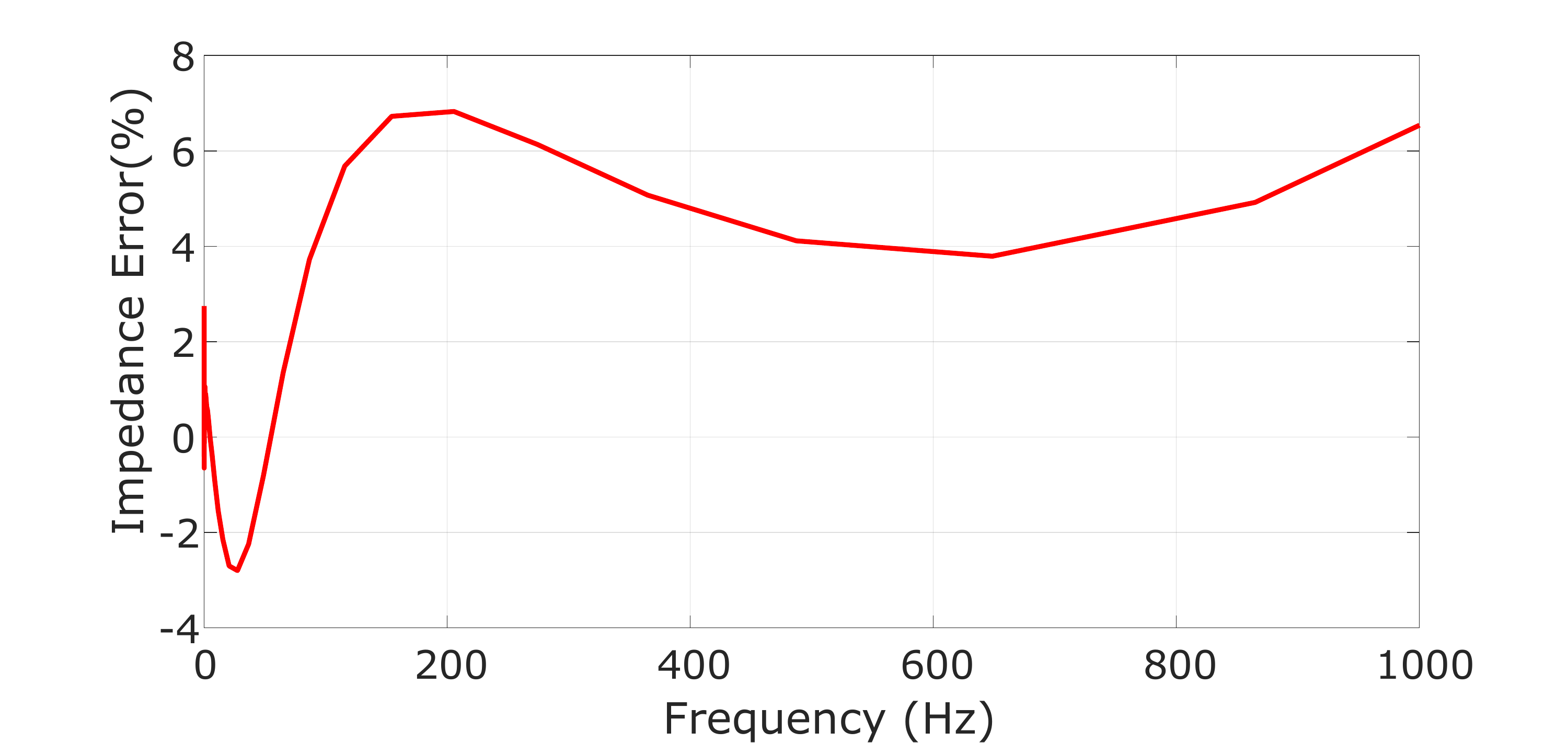}
    \caption{Impedance error plot.}
  \end{subfigure}
  \caption{Comparison of impedance plots from EIS and from the proposed ECM parameter identification method for a cell with SoC = 50\% and $T=25^\circ C$.}
  \label{fig:compare_eis_res1}
\end{figure}

\begin{figure}
  \centering
  \begin{subfigure}[b]{0.495\textwidth}
    \includegraphics[width=\textwidth]{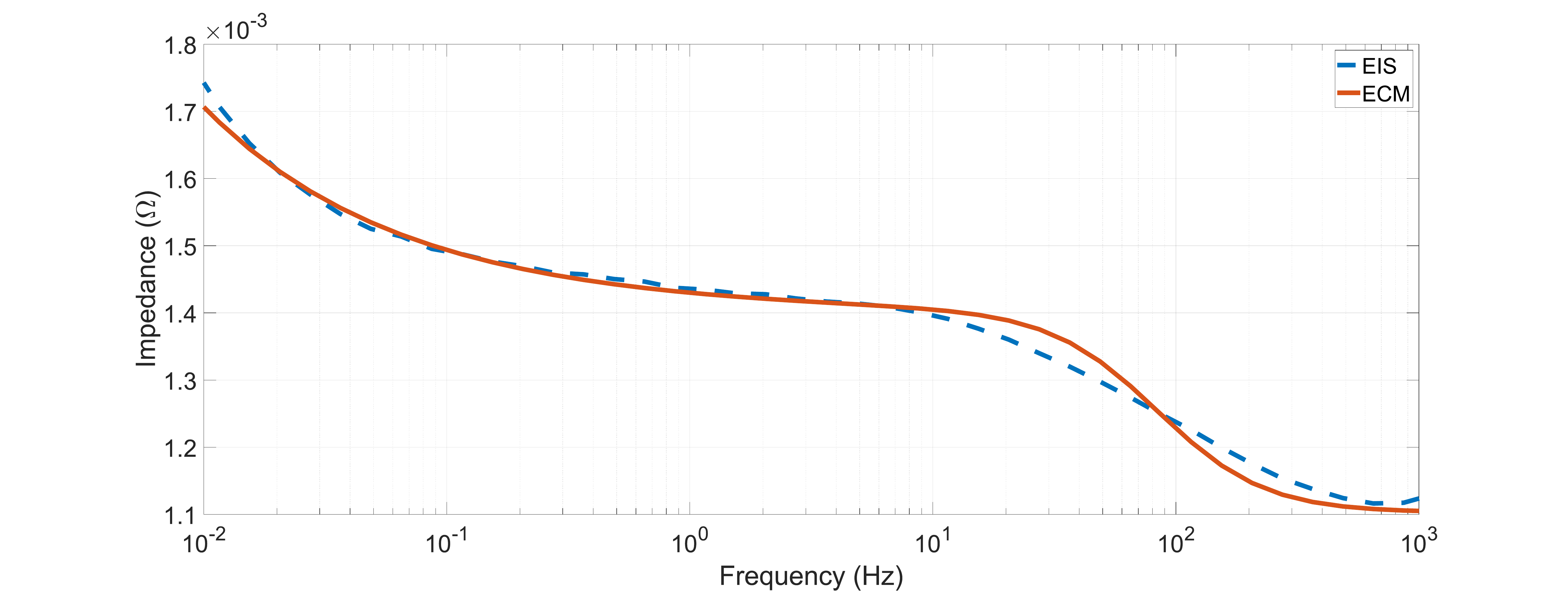}
    \caption{Impedance magnitude plot.}
  \end{subfigure}
  \hfill
  \begin{subfigure}[b]{0.495\textwidth}
    \includegraphics[width=\textwidth]{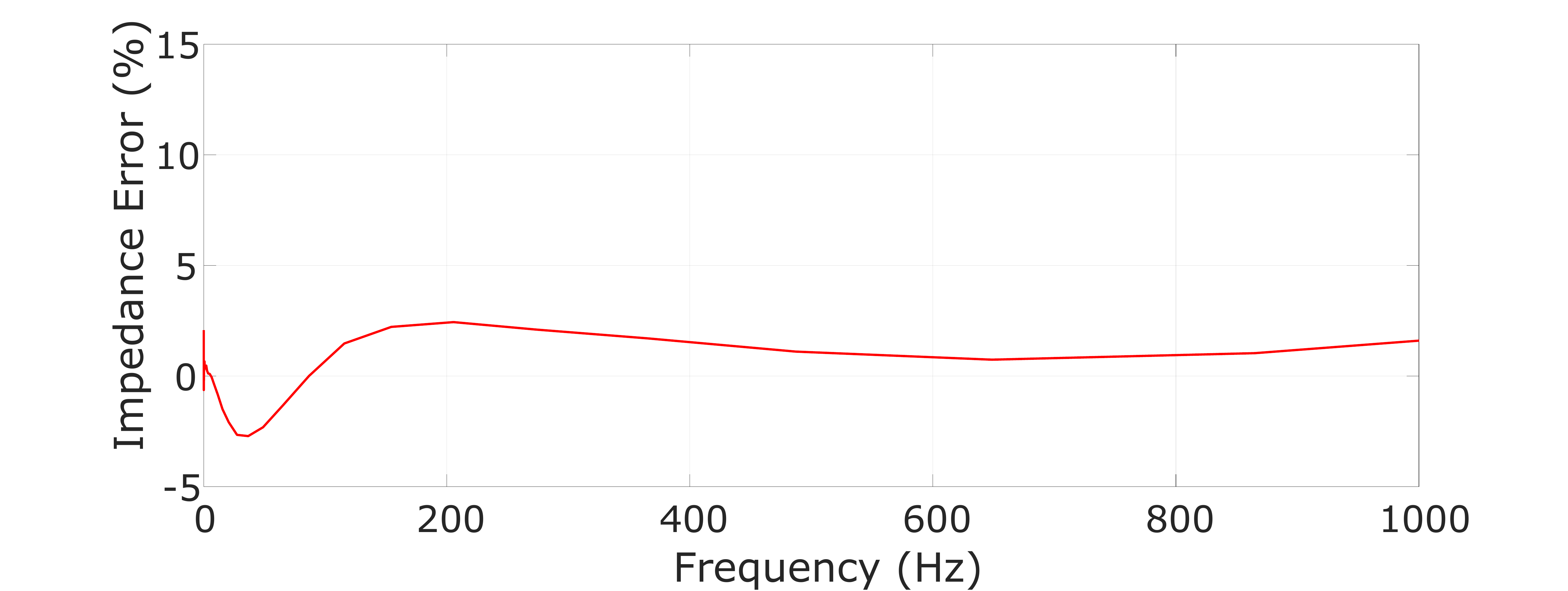}
    \caption{Impedance error plot.}
  \end{subfigure}
  \caption{Comparison of impedance plots from EIS and from the proposed ECM parameter identification method for a different cell at SoC = 40\% and $T=25^\circ C$.}
  \label{fig:compare_eis_res2}
\end{figure}

Among the impedance output of the EIS test, three discrete impedances are chosen to obtain the parameters of the ECM, which is based on Randles circuit. Fig. \ref{fig:mag_plot_freq_marked} shows impedance versus frequency for the same condition as in Fig.\ref{fig:nyquist_ex}. The discrete frequencies used for getting the proposed model parameters are also marked in this figure. That is, $\omega_{low}=0.116Hz$, $\omega_{mid}=20.55Hz$ and $\omega_{high}=648.65Hz$. 
Note that $\omega_{high}$ is chosen such that the stray inductance effect is negligible as it can be seen that the slope of the curve in Fig. \ref{fig:mag_plot_freq_marked} is zero around this point.
By following the procedure in Section 3, the ECM parameters are determined as follows.
\begin{align}
    R_0 &= 0.826m\Omega,~~R_1 = 0.346 m\Omega \notag \\
    C_1 &= 7.07F,~~~~~~A_w=0.1032m\Omega(rad/s)^{0.5} \label{eq:sample_values_ecm}
\end{align}
The parameters in Eq. \ref{eq:sample_values_ecm} are used to compare the impedance plot in Fig. \ref{fig:compare_eis_res1}.
The rms error (RMSE) in the impedance across the frequency range for this case is determined to be RMSE = 2.74\% and the peak error is 6.4\%. 
Another set of results for a \emph{different cell} at 40\% SoC and $T=25^\circ C$ is shown in Fig. \ref{fig:compare_eis_res2}. It can be observed that the results are similar to the case shown in Fig. \ref{fig:compare_eis_res2}. The RMSE for this case is 1.2\% and a peak error of 2.74\%.

\begin{figure}
    \centering
    \includegraphics[width=0.75\linewidth]{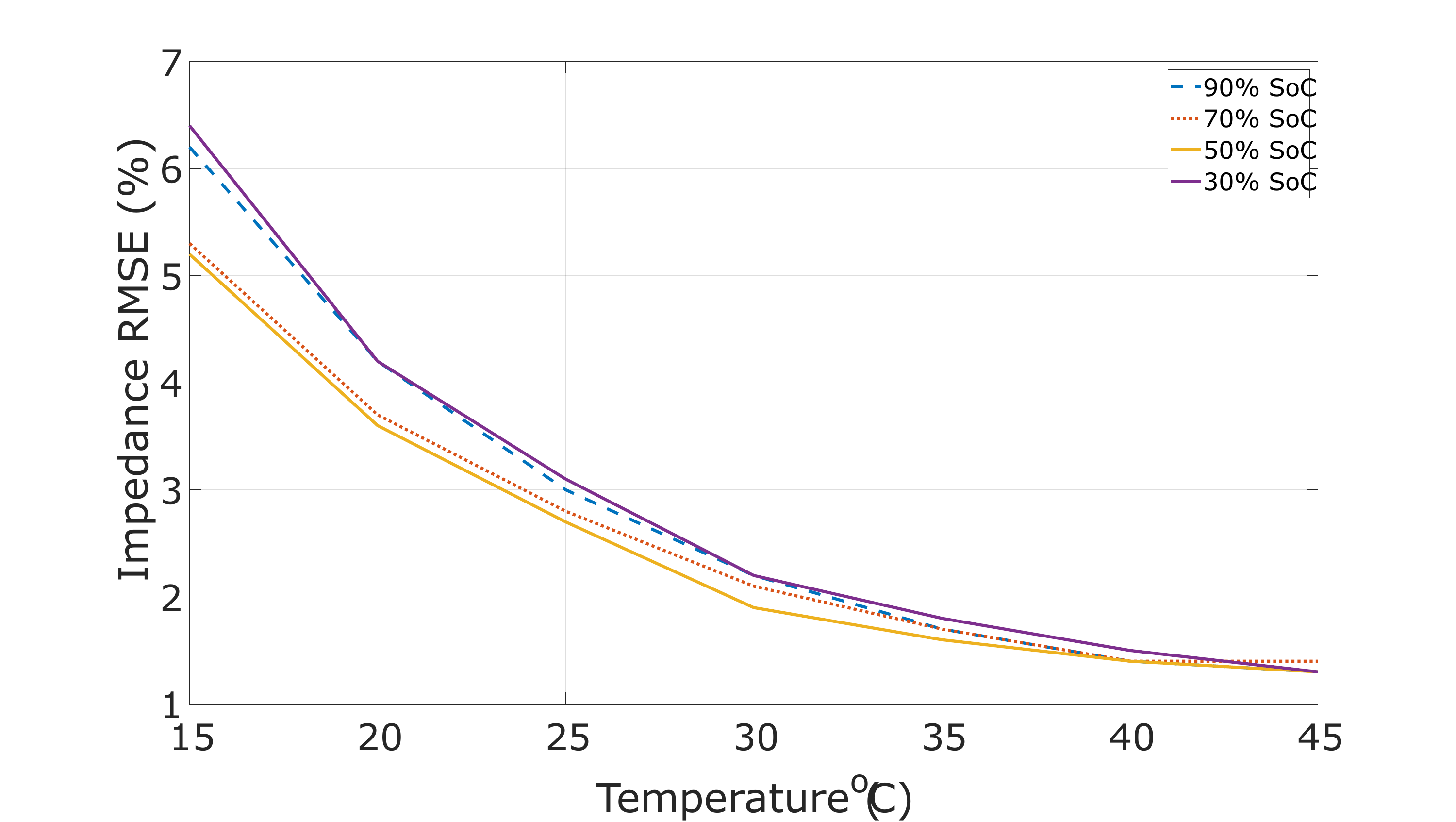}
    \caption{RMSE between the model and EIS for various conditions of ambient temperature and SoC.}
    \label{fig:error_summary1}
\end{figure}

The summary of RMSE for different temperature and SoC for a given cell are summarized in Fig. \ref{fig:error_summary1}. As it can be observed, the worst-case error is less than 6.5\%. The error reduces with the increase in ambient temperature. It can be seen that the worst case error is observed for low SoC and low temperature. In other words, if an average of the RMSE is taken across the different SoC and temperature conditions, is is observed to be $<$ 3\%. Thus, without the use of computationally intensive methods such as Fourier analysis or optimization techniques, the proposed method yields an accurate model.

\subsection{Validating the method on batteries with different capacity/geometry}
\begin{figure}
  \centering
  \begin{subfigure}[b]{0.495\textwidth}
    \includegraphics[width=\textwidth]{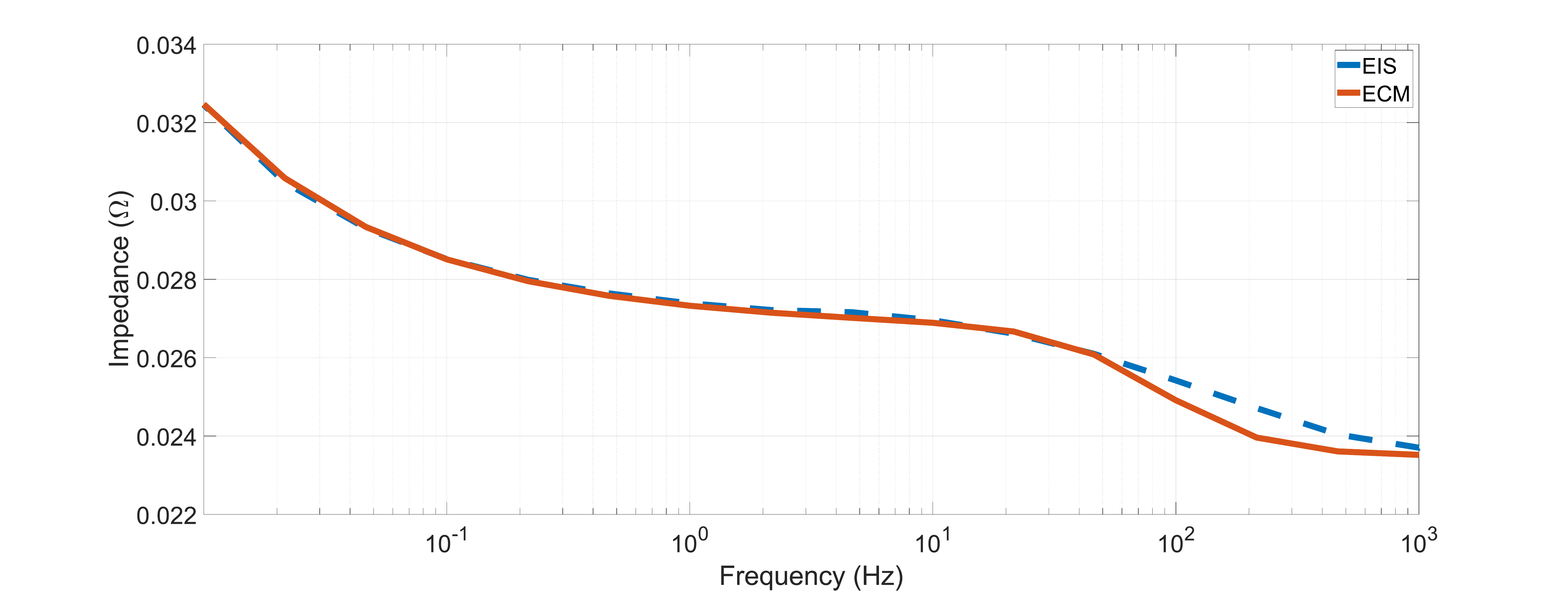}
    \caption{Impedance magnitude plot.}
  \end{subfigure}
  \hfill
  \begin{subfigure}[b]{0.495\textwidth}
    \includegraphics[width=\textwidth]{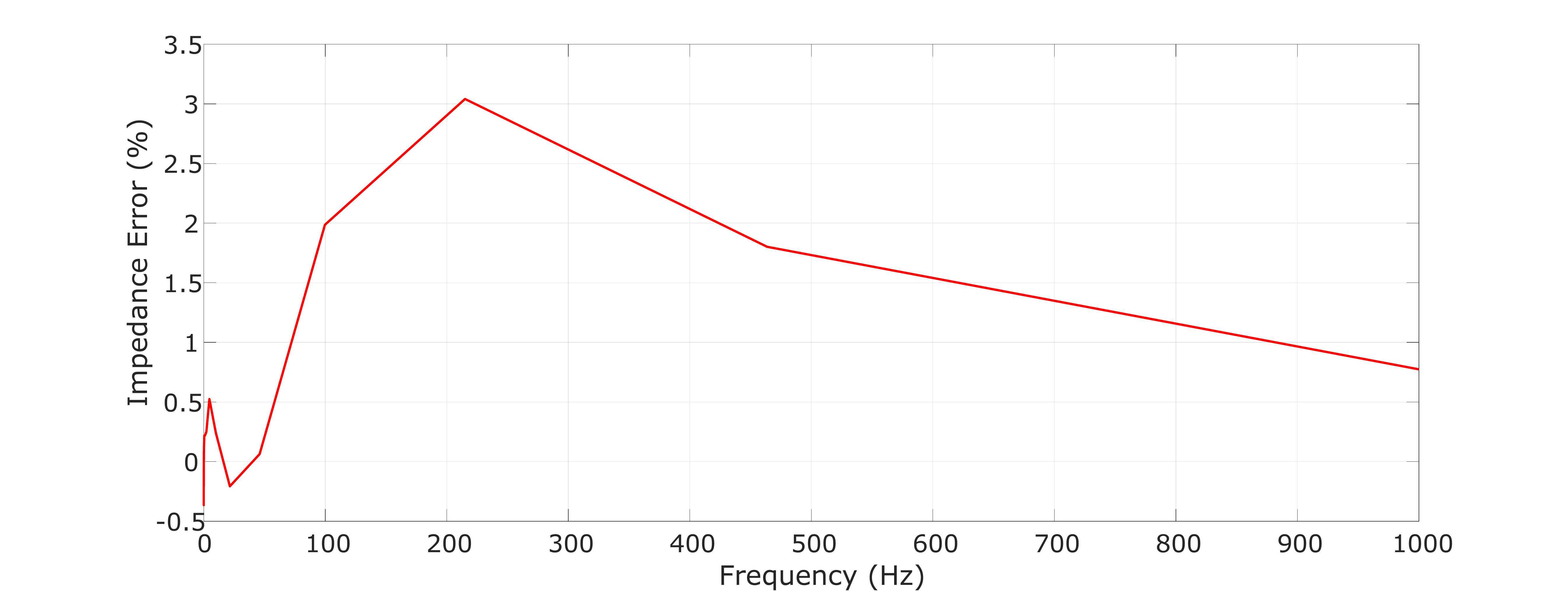}
    \caption{Impedance error plot.}
  \end{subfigure}
  \caption{Comparison of impedance plots from EIS and from the proposed ECM parameter identification method for data from Stanford University \cite{stanford_data}.}
  \label{fig:compare_eis_stanford1}
\end{figure}

The results in the Section 5.1 correspond to higher capacity prismatic cells that have very low internal impedance. To prove the versatility of the proposed approach, EIS data on cylindrical cells from Stanford University are used \cite{stanford_data}. The cells used are INR21700-M50T with graphite/silicon anode and NMC cathode. The data is available at three SoCs (20\%, 50\% and 80\%) at three different aging conditions. The aging conditions correspond to fresh cell, 151 cycles and 350 cycles \cite{stanford_data}. Fig. \ref{fig:compare_eis_stanford1}(a) shows the comparison between impedance magnitude plot from the EIS and the plot from the ECM parameters identified by proposed solution. This is for a cell at 50\% SoC which has been cycled 151 times. The corresponding errors are shown in Fig. \ref{fig:compare_eis_stanford1} (b). The RMS error is evaluated to be 1.07\%. It can also be seen that the peak error is about 3\%.

The summary of RMSE for the entire dataset along with the ECM parameters extracted is provided in Table \ref{tab:eis_stanford_full}. It can be observed that the worst-case RMSE is 1.69\%.
\begin{table}[htbp]
    \centering
    \caption{ECM parameters extracted using proposed method using the EIS dataset \cite{stanford_data}.}
    \label{tab:eis_stanford_full}
    \begin{small}   
    \begin{tabular}{ccccccc}
        \toprule
        SoC & Case & $R_0$ ($m\Omega$) & $R_1$ ($m\Omega$) & $C_1$ ($F$) & Aw ($\Omega(rad/s)^{\frac{1}{2}}$) & RMSE (\%) \\
        \midrule
         & Fresh & 22.533 & 3.352 & 0.393 & 0.001769 & 1.14 \\
      20\%  & 151 cycles & 23.658 & 5.055 & 0.604 & 0.002221 & 1.69 \\
        & 350 cycles & 24.052 & 5.457 & 0.624 & 0.002582 & 1.88 \\
        \midrule
         & Fresh & 22.421 & 2.602 & 0.354 & 0.001951 & 0.76 \\
      50\%  & 151 cycles & 23.494 & 3.317 & 0.566 & 0.001857 & 1.07 \\
        & 350 cycles & 23.841 & 3.516 & 0.631 & 0.001919 & 1.15 \\
        \midrule
         & Fresh & 22.339 & 2.472 & 0.421 & 0.002579 & 0.94 \\
      80\%  & 151 cycles & 23.358 & 3.406 & 0.666 & 0.002611 & 1.2 \\
        & 350 cycles & 23.683 & 3.879 & 0.727 & 0.002684 & 1.32 \\
        \bottomrule
    \end{tabular}
     \end{small}
\end{table}

\subsection{Validating the online impedance evaluation approach}
\begin{figure}[htbp]
    \centering
    \includegraphics[width=0.75\linewidth]{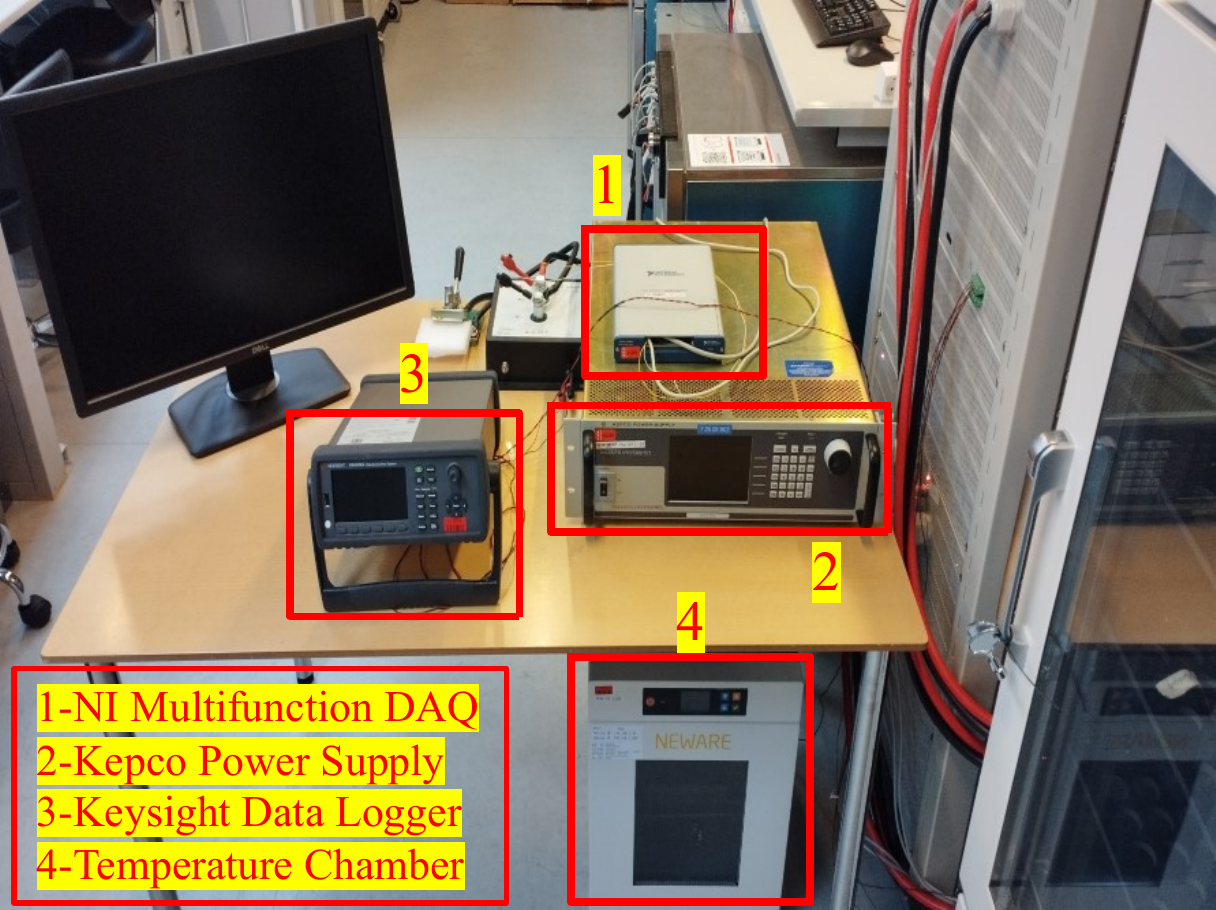}
    \caption{Experimental setup to obtain the impedance of the cell using square wave current pulses.}
    \label{fig:imp_exp_setup}
\end{figure}

\begin{figure}[htbp]
  \centering
  \begin{subfigure}[b]{0.495\textwidth}
    \includegraphics[width=\textwidth]{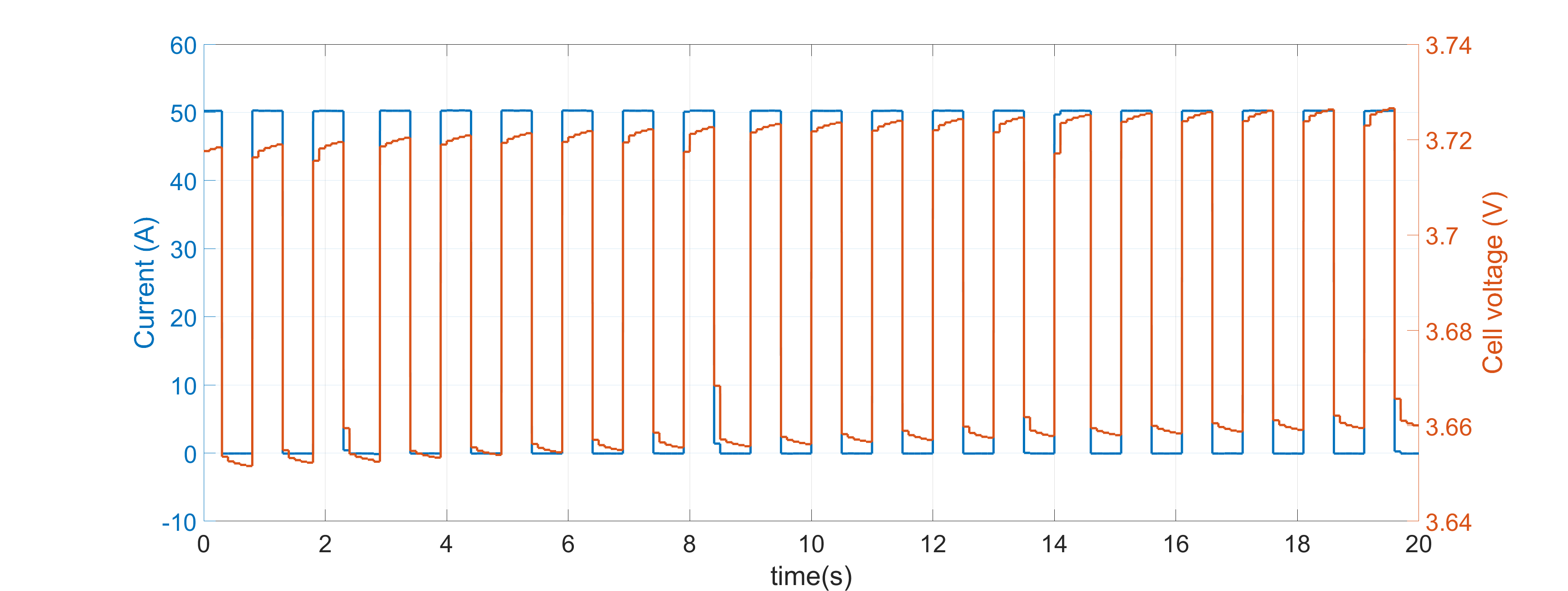}
    \caption{At $1~Hz$.}
  \end{subfigure}
  \hfill
  \begin{subfigure}[b]{0.495\textwidth}
    \includegraphics[width=\textwidth]{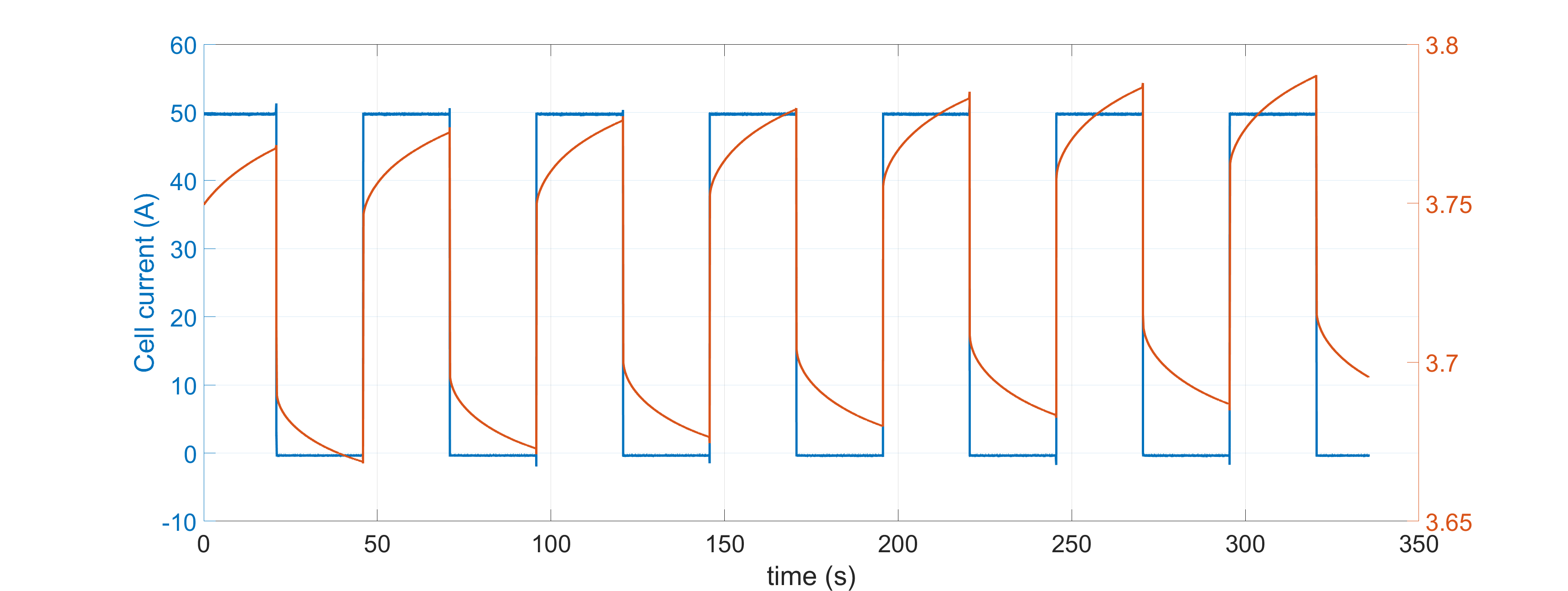}
    \caption{At $20~mHz$.}
  \end{subfigure}
  \caption{Experimental cell current and voltage using $50~A_{peak}$ pulsed charging current at 50\% duty at two frequencies.}
  \label{fig:iv_1hz_20mHz}
\end{figure}

\begin{figure}[htbp]
  \centering
  \begin{subfigure}[b]{0.49\textwidth}
    \includegraphics[width=\textwidth]{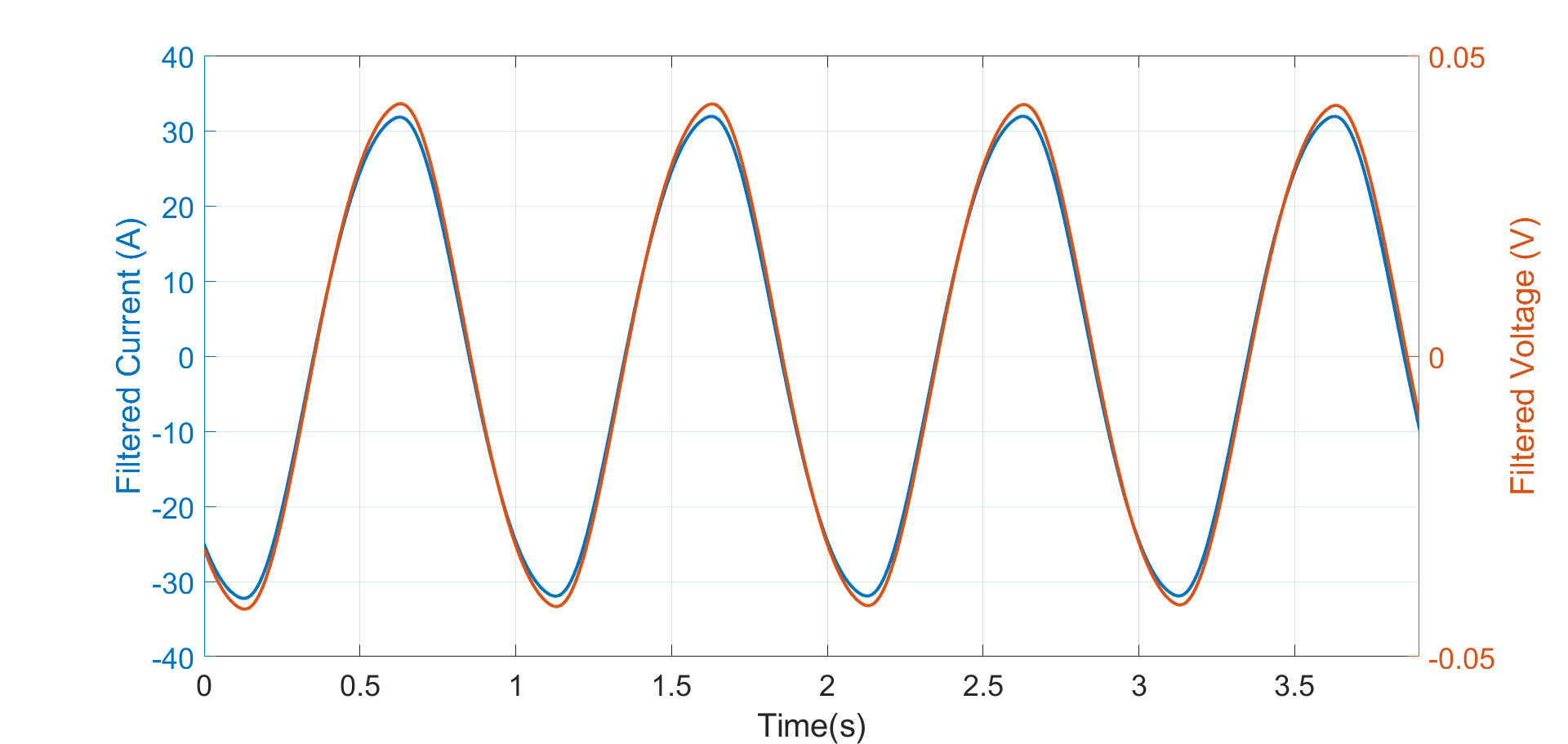}
    \caption{Fundamental voltage and current at 1Hz.}
  \end{subfigure}
  \hfill
  \begin{subfigure}[b]{0.49\textwidth}
    \includegraphics[width=\textwidth]{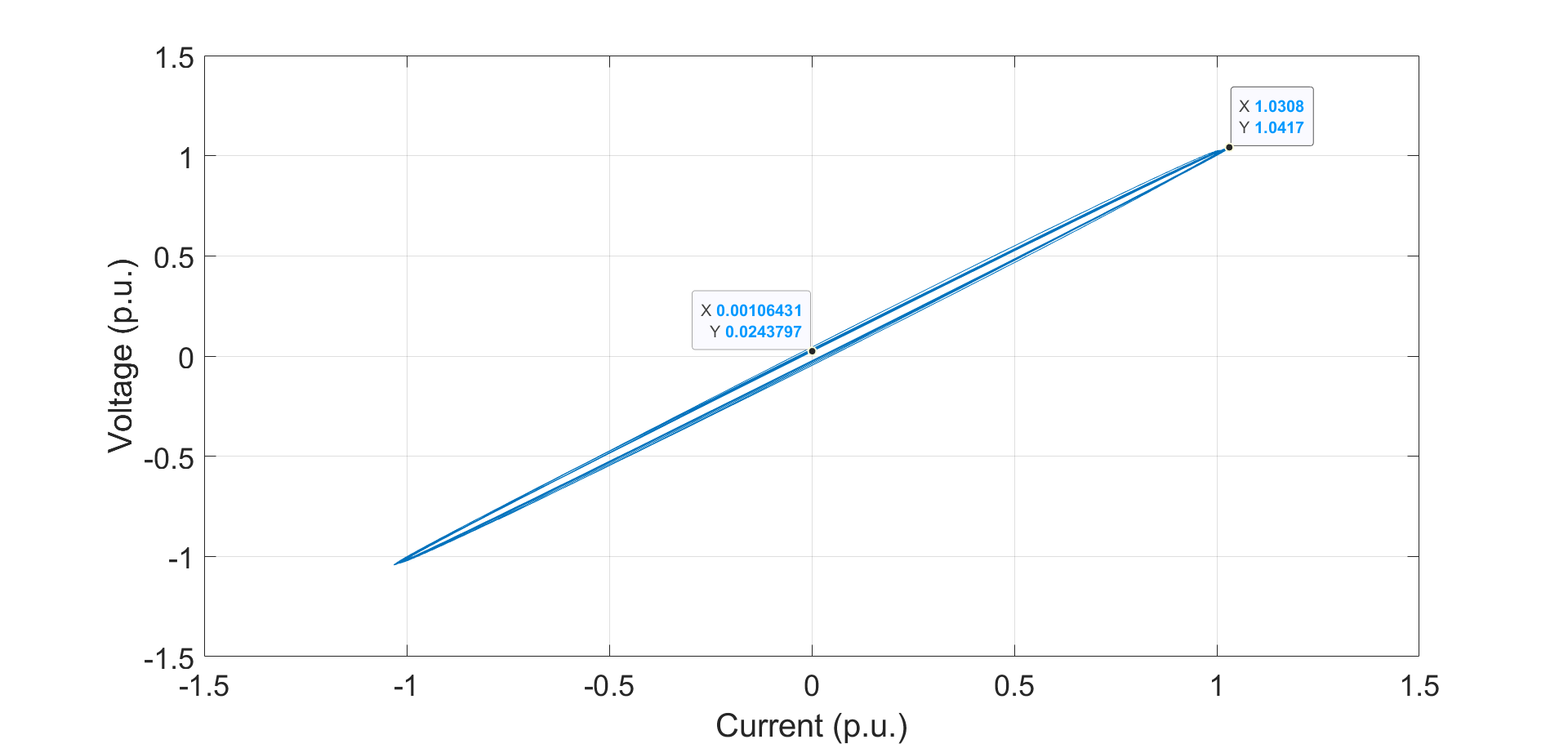}
    \caption{Lissajous pattern showing the phase difference}
  \end{subfigure}
  \caption{Impedance extraction at 1Hz.}
  \label{fig:iv_1hz_filt_combined}
\end{figure}

\begin{figure}[htbp]
  \centering
  \begin{subfigure}[b]{0.49\textwidth}
    \includegraphics[width=\textwidth]{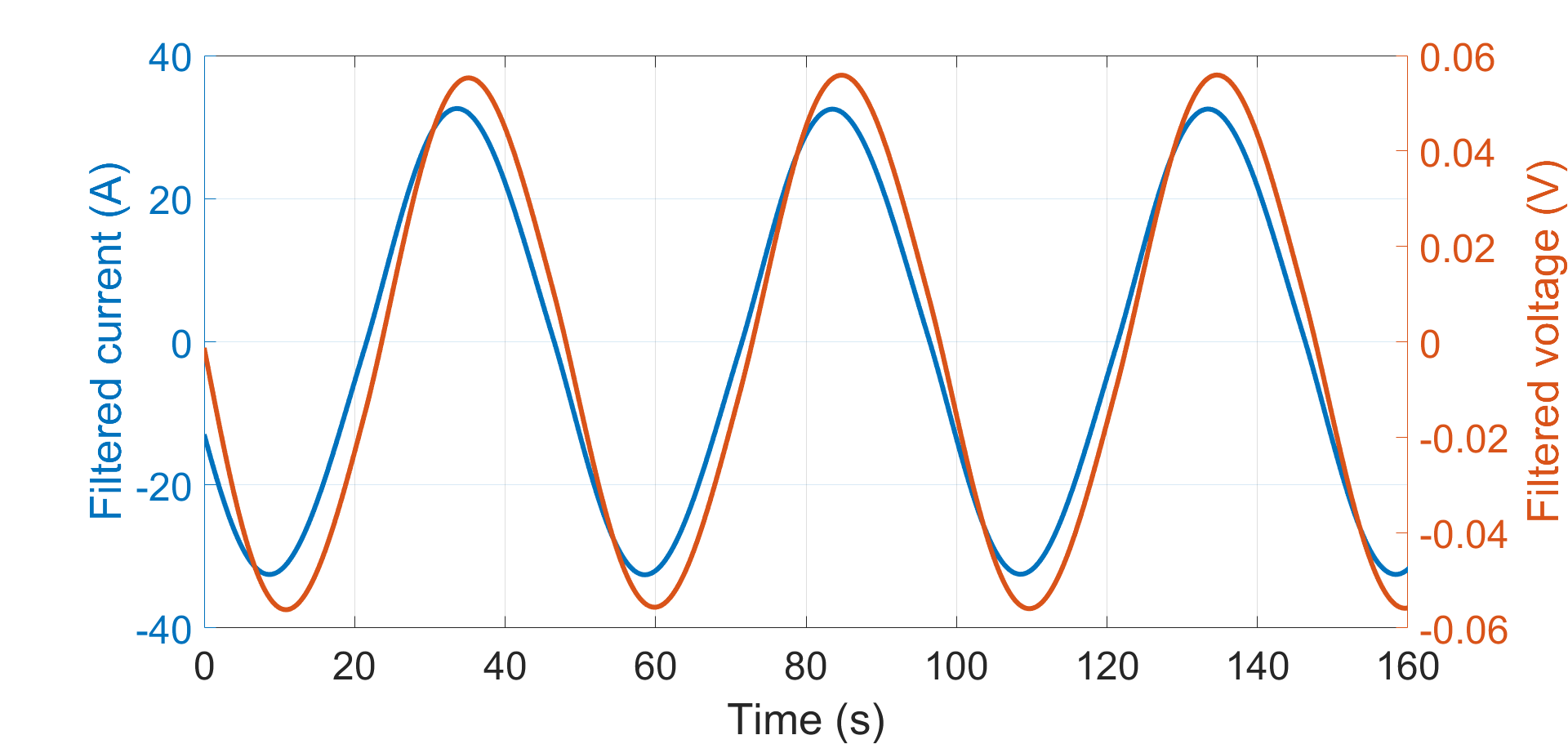}
    \caption{Fundamental voltage and current at 20mHz.}
  \end{subfigure}
  \hfill
  \begin{subfigure}[b]{0.49\textwidth}
    \includegraphics[width=\textwidth]{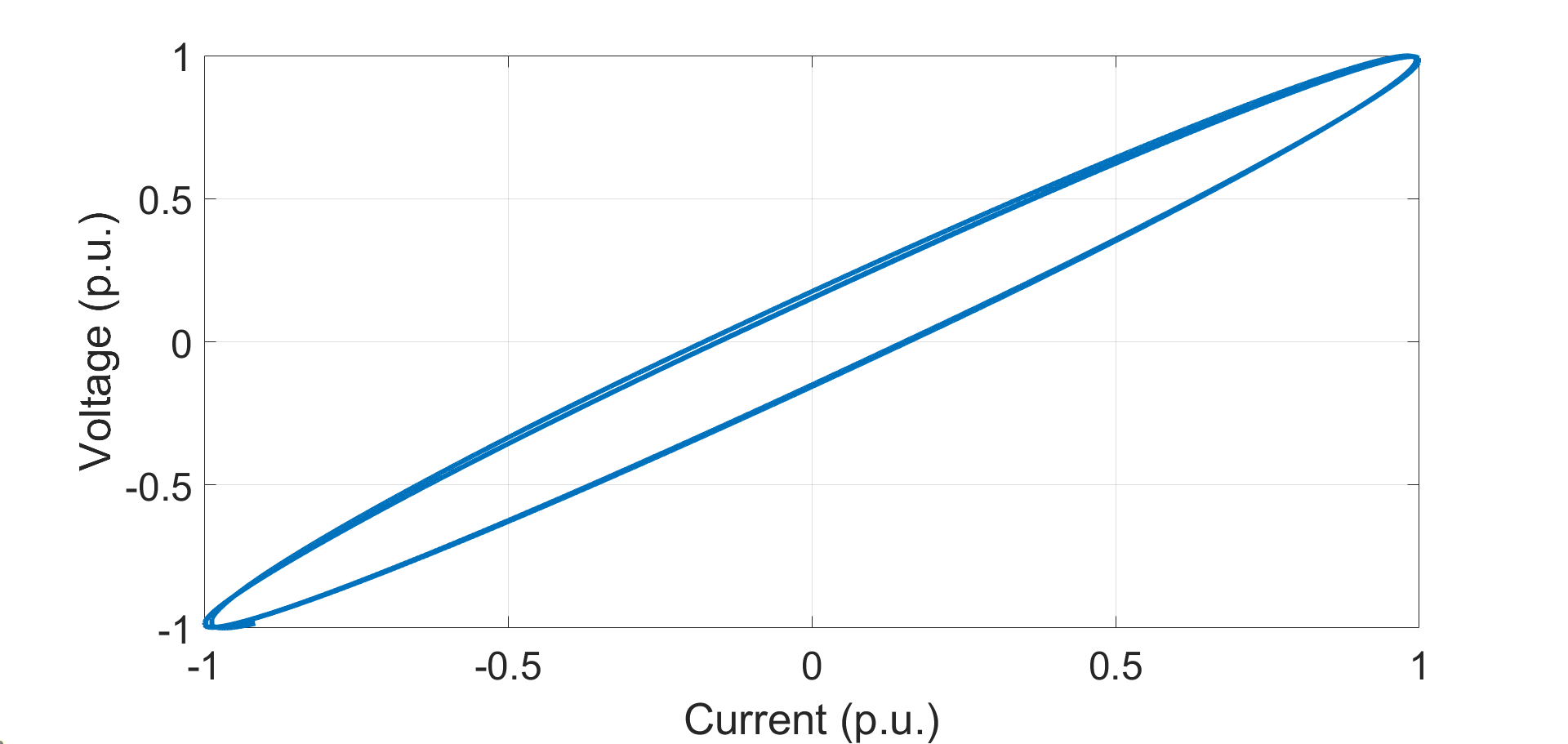}
    \caption{Lissajous pattern showing the phase difference}
  \end{subfigure}
  \caption{Impedance extraction at 20mHz.}
  \label{fig:iv_20mhz_filt_combined}
\end{figure}

An experimental setup is developed to apply the pulsed current at desired frequencies. The setup is shown in Fig. \ref{fig:imp_exp_setup}.  A bidirectional power supply Kepco is used which is controlled using Labview to produce the desired current. The cell is placed within a small thermal chamber to maintain the temperature during the test. A National Instruments (NI) data acquisition system (DAQ) is used for accurate logging of the cell voltage and current. A high resolution datalogger from Keysight is used to record the cell voltage and current. 
Fig. \ref{fig:iv_1hz_20mHz}(a) shows the cell voltage and currents for a pulse train at $1~Hz$ and Fig. \ref{fig:iv_1hz_20mHz}(b) shows for a very low frequency of $20~mHz$. 
As it can be observed, both contain the fundamental frequency component, a dc component and high-frequency harmonics. The output from a cascaded BPF with transfer function as given in Eq. \ref{eq:filt_tf}. 
The resulting fundamental voltage and current after processing from the BPF are shown in Fig. \ref{fig:iv_1hz_filt_combined}(a). 
It can be observed that there is a small phase difference between the voltage and current signals. This can be visualized using the Lissajous pattern when the per-unitized voltage is plotted against the per-unitized current to result in an elliptic waveform as shown in Fig. \ref{fig:iv_1hz_filt_combined}(b). It is well known that a straight line in this pattern corresponds to zero degree phase difference while a circular pattern corresponds to ninety degree phase difference. The phase difference in this case is observed to be $1.3\deg$ and the impedance magnitude is $1.3m\Omega$, which agrees with the data from the EIS.
Another set of results at a low frequency ($20~mHz$) where Warburg element is dominant is shown in Fig.\ref{fig:iv_20mhz_filt_combined}. It can be observed that the magnitude and phase are different for this case compared to the one at 1Hz. The phase difference is higher (about $8.3^o$) and the impedance amplitude is $1.7m\Omega$. The phase difference is again visualized using the Lissajous pattern in Fig. \ref{fig:iv_20mhz_filt_combined}(b).

\section{Conclusion}
Li ion cell ECM is a simple tool that can be used for improving the state estimation accuracy for SoC, SOH and as a digital-twin. Since the ECM parameters change considerably depending on the operating conditions and aging of the cell, this model is valuable if the BMS can perform online parameter estimation. In this paper, a novel low-complexity method is proposed which uses discrete impedance values at three frequencies to determine the ECM parameters. This method is suitable for online estimation in the BMS since it uses simple filters and algebraic equations to arrive at the model parameters. Optimization methods or curve-fitting techniques are not used and the computation is also fast.   The accuracy of the analytical method is validated experimentally by comparing the results with classical EIS and with large-signal pulses. The model parameters provide excellent accuracy for different types of cells and for a wide variation in the operating conditions in terms of SoC, temperature and SoH. The proposed methodology is shown to be suitable for online implementation. It can be used to improve the BMS accuracy in state-estimations or to enhance diagnostics and protection features for automotive applications.





\bibliographystyle{elsarticle-num} 
\biboptions{sort&compress}

\bibliography{reflist.bib}







\end{document}